\pdfoutput=1
\documentclass[nofootinbib, twocolumn, aps, prd, superscriptaddress]{revtex4-1}

\usepackage{longtable}
\usepackage{array}
\newcolumntype{C}[1]{>{\centering}m{#1}}

\usepackage{graphicx}
\usepackage{natbib}
\usepackage{color}
\newcommand{\beq}{\begin{equation}}
\newcommand{\beqa}{\begin{eqnarray}}
\newcommand{\eeq}{\end{equation}}
\newcommand{\eeqa}{\end{eqnarray}}

\newcommand{\simgt}{\lower.5ex\hbox{$\; \buildrel > \over \sim \;$}}
\newcommand{\simlt}{\lower.5ex\hbox{$\; \buildrel < \over \sim \;$}}

\newcommand{\bd}[1]{\mbox{\boldmath $#1$}}

\usepackage[usenames,dvipsnames]{xcolor}
\newcommand{\change}[1]{{\color{black} #1}}

\begin{document}
\title{
The correlation of extragalactic $\gamma$-rays
with cosmic matter density distributions \\ 
from weak-gravitational lensing
}

\author{Masato Shirasaki}
\email{masato.shirasaki@nao.ac.jp}
\affiliation{
National Astronomical Observatory of Japan (NAOJ), 
Mitaka, Tokyo 181-8588, Japan}

\author{Oscar Macias}
\email{oscar.macias@vt.edu}
\author{Shunsaku Horiuchi}
\email{horiuchi@vt.edu}
\affiliation{
Center for Neutrino Physics, Department of Physics, Virginia Tech, Blacksburg, Virginia
24061, USA
}

\author{Naoki Yoshida}
\email{naoki.yoshida@ipmu.jp}
\affiliation{
Department of Physics, University of Tokyo, Tokyo 113-0033, Japan\\
Kavli Institute for the Physics and Mathematics of the Universe (WPI),
University of Tokyo, Kashiwa, Chiba 277-8583, Japan
}
\affiliation{
CREST, Japan Science and Technology Agency, 4-1-8 Honcho, Kawaguchi, Saitama, 332-0012, Japan
}

\author{Chien-Hsiu Lee}
\affiliation{
Subaru Telescope, NAOJ, 650 N Aohoku Pl. Hilo, HI 96720, USA
}

\author{Atsushi J. Nishizawa
}
\affiliation{Institute for Advanced Research, Nagoya University, Nagoya 464-8602, Aichi, Japan}

\begin{abstract}
The extragalactic $\gamma$-ray background (EGB) arises from the accumulation of $\gamma$-ray emissions from resolved and unresolved extragalactic sources as well as diffuse processes. 
It is important to study the statistical properties of the EGB in the context of cosmological structure formation.
Known astrophysical $\gamma$-ray sources such as blazars, star-forming galaxies, and radio galaxies are expected to trace the underlying cosmic matter density distribution.
We explore the correlation of the EGB from Fermi-LAT data with the large-scale matter density distribution from the Subaru Hyper Suprime-Cam (HSC) SSP survey.
We reconstruct an unbiased surface matter density distribution $\kappa$ at $z\simlt 1$ by applying weak-gravitational lensing analysis to the first-year HSC data. We then calculate the $\gamma - \kappa$ cross-correlation.
\change{Our measurements are consistent with a null detection, but a weak correlation 
is found at angular scales of 30-60 arcmin, especially when distant source galaxies 
at $z > 1$ are used for the lensing $\kappa$ reconstruction. 
The large-scale correlation suggests strong clustering of
high-redshift $\gamma$-ray sources such as blazars. However,
the inferred bias factor of $4-5$ is larger by about a factor of two
than results from other clustering analyses. 
The final HSC data covering 1,400 squared degrees will play an essential role 
to determine accurately the blazar bias at $z > 0.5$.
}
\end{abstract}

\maketitle

\section{\label{sec:intro}INTRODUCTION}

Since its discovery, 
extragalactic $\gamma$ rays have been one of the most important subjects 
in high-energy astrophysics \cite{1972ApJ...177..341K}. The extragalactic gamma-ray background (EGB)  is partly explained by $\gamma$-ray emission from resolved point sources outside the Milky Way. The Large Area Telescope (LAT) onboard the Fermi Gamma-ray Space Telescope has detected $\gamma$-ray sources from across the whole sky \cite{Acero:2015hja}. Many of them are extragalactic, and hence contribute to the EGB. In addition to the resolved contributions, a diffuse and nearly isotropic emission, referred to as the isotropic gamma-ray background (IGRB), is necessary 
to provide a better fit to the EGB observations \cite{Sreekumar:1997un, Ackermann:2014usa} (also see Ref.~\cite{Fornasa:2015qua} for a recent review of the IGRB). Although the origin of the IGRB is still under debate, a significant fraction may originate from the emission from unresolved astrophysical sources, which are too faint to be detected on an individual basis. Possible candidates include blazars \cite{Collaboration:2010gqa, Harding:2012gk},
star-forming galaxies \cite{Thompson:2006qd, Makiya:2010zt}, and radio galaxies \cite{Inoue:2011bm, DiMauro:2013xta}.

Recently, Ref.~\cite{Ackermann:2014usa} measured the energy spectrum of the EGB in the range between 0.1 GeV to 820 GeV and showed that the EGB spectrum is well described by a power law with a photon index of $2.32\, (\pm 0.02)$ and an exponential cut off of $279\, (\pm52)$ GeV. Improved modeling of a linear combination of astrophysical $\gamma$-ray sources based on Fermi-LAT data can provide an adequate interpretation of the origin of the EGB. Ref.~\cite{Ajello:2015mfa} showed that the cumulative emission of blazars, star-forming galaxies, and radio galaxies can explain the EGB spectrum in the range 0.1--820 GeV with the cut off originating from the expected $\gamma$-ray attenuation by the extragalactic background light \cite{Gould:1966pza}.

In the standard model  of cosmic structure formation, galactic-size astrophysical sources are
hosted by dark matter halos. Therefore, the spatial distribution of EGB emission is expected to be correlated with the underlying matter density distribution of the Universe \cite{Camera:2012cj, Fornengo:2013rga}. There already exist several pieces of supporting evidence that show such a correlation between the EGB and 
large-scale structure (LSS) tracers. For example, Ref.~\cite{Allevato:2014qga} find spatial clustering of resolved blazars detected in the 2-year all-sky survey by Fermi-LAT. The cross correlations between IGRB and galaxies \cite{Xia:2015wka} and galaxy clusters \cite{Branchini:2016glc} have been also detected. However, these previous studies do not provide  direct information on the relation between the EGB and the cosmic matter density, because the tracers of LSS that have been used are known to be {\it biased} with respect to the underlying matter density.

Fortunately, gravitational lensing provides a physical and {\it unbiased} probe of the cosmic matter density \cite{Bartelmann:1999yn}. It is free from assumptions that relate the matter density to observable quantities. By contrast, e.g., for galaxies to be used as a tracer of large-scale structure, some assumption needs to be made on the relation between the galaxy luminosity and mass, and even additional conditions need to be invoked such as hydrostatic equilibrium. 

The correlation between the IGRB and large-scale
matter distribution has been detected using gravitational lensing of the cosmic microwave background (CMB), reported in Ref.~\cite{Fornengo:2014cya}. In the concordance $\Lambda$CDM cosmology, the CMB lensing effect is largely 
determined by the matter density distribution at redshift $z=2-3$ \cite{Lewis:2006fu}. Thus, it is still unclear whether there exists a strong correlation between the EGB and cosmic matter density at $z\simlt 1$. 
Since extragalactic sources at lower redshift should appear brighter, 
a large fraction of the EGB is expected to originate 
from the structures (sources) at $z\simlt 1$. This fact further motivates to use cross-correlation analysis of the EGB with low-redshift structures.

In this paper, we explore the correlation between the EGB and the cosmic matter density distribution probed by an optical weak lensing survey. We achieve this by using galaxy imaging data taken from the Subaru Hyper-Suprime Cam (HSC) SSP survey. Unlike similar investigations in the literature \cite{Shirasaki:2014noa, Shirasaki:2016kol, Troster:2016sgf},
we work with the reconstructed surface matter density from the galaxy imaging data, i.e., lensing convergence. We then perform, for the first time, the cross-correlation analysis of the EGB (including all the resolved components) and the lensing convergence. Our measurement can provide direct evidence of the correlation between the extragalactic $\gamma$-rays
and cosmic matter density at $z\simlt1$. 
It also enables us to study the {\it clustering} properties of the EGB, and provides invaluable information on the EGB sources that cannot be extracted from mean intensity analyses.
We expect that the correlation depends on the clustering of extragalactic $\gamma$-ray sources at redshifts of $z\sim0.5$.
Since it is difficult to measure the auto-correlation of $\gamma$-ray sources 
as a function of redshift because of the significantly low number density of known  sources, our measurement will be complementary to other analyses and give important implications 
for 
the nature of the EGB.

The paper is organized as follows. 
In Section~\ref{sec:obs}, 
we summarize the basics of EGB
and gravitational lensing. 
In Section~\ref{sec:data}, we describe 
the galaxy-imaging and $\gamma$-ray data used, 
and provide details of the cross-correlation analysis. 
Our benchmark model of
the cross correlation is discussed in Section~\ref{sec:model}. 
In Section~\ref{sec:res}, we show the result of 
our cross-correlation analysis, and 
discuss constraints on possible connection between astrophysical
$\gamma$-ray sources and LSS.  Concluding remarks and discussions are given in Section~\ref{sec:con}. 
Throughout, we use the standard cosmological parameters
$H_0=100h\, {\rm km}\, {\rm s}^{-1}$
with $h=0.7$, the average matter density 
$\Omega_{\rm m0}=0.279$, the cosmological constant
$\Omega_{\Lambda}=0.721$,
and the amplitude of matter density fluctuations within 
$8\,h^{-1}\, {\rm Mpc}$, $\sigma_8=0.821$.

\section{OBSERVABLES}
\label{sec:obs}

\subsection{Extragalactic gamma-ray background}
\label{subsec:EGB}

The $\gamma$-ray intensity along a given direction $\bd{\theta}$
can be written as \cite{Fornengo:2013rga},
\beqa
I_{\gamma}(\bd{\theta})=\int {\rm d}\chi \, W_{g}(\chi)g(r(\chi)\bd{\theta},\chi),
\label{eq:gamma_intensity}
\eeqa
where 
$\chi(z)$ 
denotes the radial comoving distance (function of redshift $z$), 
$r(\chi)$ is the comoving angular diameter distance,
$g(\bd{x})$ is the relevant density field of the $\gamma$-ray source,
and
$W_{g}(\chi)$ is the window function.
In this paper, we define $W_{g}(\chi)$ so that the mean $\gamma$-ray intensity can be expressed as,
\beqa
\langle I_{\gamma} \rangle = \int {\rm d}\chi\, W_g(\chi),
\label{eq:mean_gamma_intensity}
\eeqa
where $\langle \cdots \rangle$ represents an averaged quantity
over the sky.

Suppose that an astrophysical source population $X$
can emit $\gamma$ rays, we set the relevant field in Eq.~(\ref{eq:gamma_intensity}) to be
$g(\bd{x})\propto L_{\gamma}\delta^{(3)}(\bd{x})$ 
where $L_{\gamma}$ denotes the $\gamma$-ray intensity emitted from sources,
$\delta^{(3)}(\bd{x})$ is the dirac delta function in three-dimensional space and we assume any source populations can be approximated as point source\footnote{This approximation
should be valid as long as we focus on the correlation with the scale much larger than the actual size of astrophysical sources.
The angular scales of interest in this paper is set to be larger than 30 arcmins.}.
In terms of the luminosity function, the window function $W_g$
for the source population $X$
can be written as \cite{Camera:2012cj, Camera:2014rja},
\beqa
W_{g, X}(\chi) &=&
\int_{E_{\rm min}}^{E_{\rm max}} \frac{{\rm d}E_{\gamma}}{4\pi}\,
e^{-\tau\left(E_{\gamma}^{\prime},z(\chi)\right)} 
\nonumber \\
&&
\,\,\,\,\,
\times
\int_{\Gamma_{\rm min}}^{\Gamma_{\rm max}} {\rm d}\Gamma_{X}\, {\cal A}(z(\chi), \Gamma_X)
\frac{{\rm d}N_{\gamma,X}(E_{\gamma}, z(\chi), \Gamma_{X})}{{\rm d}E_{\gamma}}
\nonumber \\ 
&&
\,\,\,\,\,
\times
\int_{L_{\rm min}}^{L_{\rm max}} {\rm d}L_{\gamma}\,
L_{\gamma}\Phi_{X}(L_{\gamma}, z(\chi), \Gamma_X),
\label{eq:EGB_window}
\eeqa
where
$E_\gamma$ is the observed $\gamma$-ray energy, 
$E^{\prime}_\gamma = (1+z) E_\gamma$ is the energy 
of the $\gamma$ ray at redshift $z$,
${\rm d}N_{\gamma, X} /{\rm d}E_\gamma$ 
is the $\gamma$-ray spectrum,
$\Gamma_{X}$ is the photon spectrum index,
$\Phi_{X}(L_\gamma, z, \Gamma_{X})$ is 
the $\gamma$-ray luminosity function,
and the exponential factor in the integral
takes into account the effect of $\gamma$-ray attenuation 
during propagation owing to pair creation 
on diffuse extragalactic photons. 
For the $\gamma$-ray optical depth 
$\tau\left(E^{\prime}_\gamma, z \right)$, 
we adopt the model in Ref.~\cite{Gilmore:2011ks}.
The factor ${\cal A}$ in Eq.~(\ref{eq:EGB_window}) 
accounts for the conversion normalisation between the emitted and observed 
$\gamma$-ray energies including k-correction, and depends on the
source spectral index and the energy range considered.

In the following, we summarize the basic quantities to compute
the $\gamma$-ray intensity using Eq.~(\ref{eq:EGB_window})
for several astrophysical $\gamma$-ray sources.

\subsubsection{Blazars}

Following Ref.~\cite{Ajello:2015mfa},
we characterize the blazar population by means of a parametric description of their $\gamma$-ray luminosity function and  
energy spectrum.
The blazer $\gamma$-ray luminosity function
is defined as the number of sources per unit luminosity
$L_{\gamma}$ (defined in the rest frame of the source, for
energies between 0.1 and 100 GeV). 
The luminosity function at $z=0$ is modeled as, 
\beqa
\Phi_{b}(L_{\gamma},z=0,\Gamma_b)
&=& \frac{A}{\ln 10L_{\gamma}}\left[
\left(\frac{L_{\gamma}}{L_{\ast}}\right)^{\gamma_a}
+\left(\frac{L_{\gamma}}{L_{\ast}}\right)^{\gamma_b}\right]^{-1}
\nonumber \\
&&
\,\,\,\,\,
\,\,\,\,\,
\times
e^{-\left[\Gamma_{b}-\mu(L_{\gamma})\right]^2/2\sigma^2},
\label{eq:blazar_lf}
\eeqa
where the parameters of $A$, 
$\gamma_a$, $\gamma_{b}$, $L_{\ast}$, $\sigma$
and the relation of 
$\mu(L_{\gamma})$ have been calibrated with 
Fermi-LAT resolved blazars for 
a given redshift-evolution scenario \cite{Ajello:2015mfa}.
We assume that the redshift evolution 
of the luminosity function is expressed in the following form,
\beqa
\Phi_{b}(L_{\gamma},z,\Gamma_b)&=&f_e(z,L_\gamma)\Phi_{b}(L_{\gamma},z=0,\Gamma_b),\\ \label{eq:LDDE}
f_e(z,L_{\gamma})&=&
\Biggl[
\left(\frac{1+z}{1+z_c(L_{\gamma})}\right)^{-p_1(L_{\gamma})}
\nonumber \\ 
&&
+\left(\frac{1+z}{1+z_c(L_{\gamma})}\right)^{-p_2(L_{\gamma})}
\Biggr]^{-1},
\eeqa
where $z_c$, $p_1$ and $p_2$ are assumed to depend on 
the $\gamma$-ray luminosity.
We use the same parameter values as in Ref.~\cite{Ajello:2015mfa}
in Eqs.~(\ref{eq:blazar_lf}) and (\ref{eq:LDDE}). This model
is called the luminosity-density dependent evolution scenario.  

Blazars are known to have a characteristic, broad energy spectrum.
We follow the $\gamma$-ray energy spectrum proposed in
Ref.~\cite{Ajello:2015mfa},
\beqa
\frac{{\rm d}N_{\gamma, b}}{{\rm d}E_{\gamma}}
= K \left[
\left(\frac{E_{\gamma}}{E_b(\Gamma_b)}\right)^a
+
\left(\frac{E_{\gamma}}{E_b(\Gamma_b)}\right)^b
\right]^{-1},
\eeqa
where $\log (E_b/1\, {\rm GeV}) = 9.25-4.11\Gamma_b$
and we determine the normalization factor $K$ 
so that $L_{\gamma}$ 
can be defined in the energy range of 0.1-100 GeV.
We set ${\cal A} = (1+z)^{-\Gamma_b}$ to include the k-correction.
When computing the $\gamma$-ray intensity from blazars,
we adopt $\Gamma_{\rm min}=1$, $\Gamma_{\rm max}=3.5$,
$L_{\rm min}=10^{43}\, {\rm erg}\, {\rm s}^{-1}$,
and $L_{\rm max}=10^{52}\, {\rm erg}\, {\rm s}^{-1}$.

Note that the above model of $\Phi_{b}(L_{\gamma},z,\Gamma_b)$
and ${\rm d}N_{\gamma, b}/{\rm d}E_{\gamma}$
can reasonably explain the Fermi-LAT data of resolved blazars.
The expected contribution to the mean EGB intensity 
above $\sim1$ GeV will be $\sim60-70$\% \cite{Ajello:2015mfa}.

\subsubsection{Star-forming galaxies}

For star-forming galaxies, 
we assume the power-law energy spectrum  
${\rm d}N_{\gamma,s}/{\rm d}E_{\gamma}\propto E^{-2.2}_{\gamma}$,
characteristic of the LAT-detected starburst galaxies \cite{Ackermann:2012vca}. 
We obtain the luminosity function by scaling from the 
infrared luminosity function of galaxies measured in 
Ref.~\cite{Rodighiero:2009up}.
The conversion of the infrared luminosity to the 
gamma-ray luminosity of the galaxies is given by
\beqa
\log\left(\frac{L_\gamma}{{\rm erg}\, {\rm s}^{-1}}\right)
= 1.17 \log\left(\frac{L_{\rm IR}}{10^{10}L_{\odot}}\right)+39.28,
\eeqa
where $L_\gamma$ is defined in the range of 0.1-100 GeV in observer's frame
and $L_{\rm IR}$ is the infrared luminosity for 8-1000 $\mu$m \cite{Ackermann:2012vca}.
According to the definition of $L_\gamma$, we set ${\cal A}=(1+z)^{-2}$.
We adopt $L_{\rm min}=10^{30}\, {\rm erg}\, {\rm s}^{-1}$
and $L_{\rm max}=10^{43}\, {\rm erg}\, {\rm s}^{-1}$.
  
\subsubsection{Radio galaxies}

For radio galaxies,
we follow the model of Ref.~\cite{DiMauro:2013xta},
which has established a correlation
between the $\gamma$-ray luminosity and 
the radio-core luminosity $L_{r,\rm core}$ at 5 GHz. 
Using the correlation together with the radio luminosity 
function \cite{Willott:2000dh}, one can evaluate the  
contribution to the EGB from radio galaxies. 
We consider the best-fit $L_{\gamma}-L_{r,\rm core}$ relation
from Ref.~\cite{DiMauro:2013xta},
\beqa
\log\left(\frac{L_\gamma}{{\rm erg}\, {\rm s}^{-1}}\right)
= 1.008 \log\left(\frac{L_{r, \rm core}}{{\rm erg}\, {\rm s}^{-1}}\right)+2.00, \label{eq:Lgamma_to_Lradio}
\eeqa
where 
$L_{\gamma}$ is defined between 0.1 and 100 GeV in observer's frame.
We assume an average spectral index of 2.37 throughout this paper,
and set ${\cal A}=(1+z)^{-2}$.
$L_{\rm min}=10^{41}\, {\rm erg}\, {\rm s}^{-1}$
and $L_{\rm max}=10^{49}\, {\rm erg}\, {\rm s}^{-1}$ are adopted
in Eq.~(\ref{eq:EGB_window}).

\subsection{Gravitational lensing}

\subsubsection{Basics}

The weak gravitational lensing effect is commonly characterized by
the distortion of image of a source object (galaxy) by the following $2\times2$ matrix,
\beqa
A_{ij} = \frac{\partial x_{\rm true}^{i}}{\partial x_{\rm obs}^{j}}
           \equiv \left(
\begin{array}{cc}
1-\kappa -\gamma_{1} & -\gamma_{2}-\omega  \\
-\gamma_{2}+\omega & 1-\kappa+\gamma_{1} \\
\end{array}
\right), \label{eq:distortion_tensor}
\eeqa
where we denote the observed position of a source object as $\bd{x}_{\rm obs}$ 
and the true position as $\bd{x}_{\rm true}$.
In the above equation, $\kappa$ is the convergence, $\gamma$ is the shear, and $\omega$ is the rotation.
In the weak lensing regime ($\kappa, \gamma \ll 1$), one can relate the convergence
field to the density contrast of underlying matter density field $\delta_m(\bd{x})$ \cite{Bartelmann:1999yn},
\beqa
\kappa(\bd{\theta}) &=& \int_{0}^{\infty}{\rm d}\chi\, W_{\kappa}(\chi) \delta_m(r(\chi)\bd{\theta}, \chi), \\ \label{eq:delta2kappa}
W_{\kappa}(\chi) &=& \frac{3}{2}\left(\frac{H_{0}}{c}\right)^2\Omega_{\rm m0}(1+z(\chi))r(\chi) \nonumber \\
&&
\,\,\,\,\,
\,\,\,\,\,
\,\,\,\,\,
\,\,\,\,\,
\times
\int_{\chi}^{\infty}{\rm d}\chi^{\prime}\,
p(\chi^{\prime})\frac{r(\chi^{\prime}-\chi)}{r(\chi^{\prime})},
\label{eq:lens_kernel}
\eeqa
where 
$p(\chi)$ represents 
the source distribution normalized to 
$\int{\rm d}\chi\,p(\chi)=1$.

\subsubsection{Reconstruction of convergence field}
\label{subsec:reconst_kappa}

In optical imaging surveys, galaxies' ellipticities are commonly used to estimate the shear component $\gamma$ in Eq.~(\ref{eq:distortion_tensor}).
Since each component in the tensor $A_{ij}$
can be expressed as the second derivative of the gravitational potential, one can reconstruct the convergence field 
from the observed shear field in Fourier space as 
\beqa
\hat{\kappa}(\bd{\ell})
=\frac{\ell_{1}^2-\ell_{2}^2}{\ell_{1}^2+\ell_{2}^{2}}\hat{\gamma}_{1}(\bd{\ell})
+\frac{2\ell_{1}\ell_{2}}{\ell_{1}^2+\ell_{2}^{2}}
\hat{\gamma}_{2}(\bd{\ell}),
\label{eq:gamma2kappa}
\eeqa
where $\hat{\kappa}$ and $\hat{\gamma}$
are the convergence and shear in Fourier space, 
and $\bd{\ell}$ is the wave vector with components $\ell_{1}$
and $\ell_{2}$ \cite{Kaiser:1992ps}.

For a given source galaxy, one considers 
the relation between the observed ellipticity 
$e_{{\rm obs},\alpha}$ and the expected shear 
$\tilde{\gamma}_{\alpha}$,
\beqa
\tilde{\gamma}_{\alpha} 
&=& \frac{e_{{\rm obs},\alpha}}{2\cal{R}} ,\\
\tilde{\gamma}_{\alpha}
&=& (1+m) \gamma_{{\rm true}, \alpha} + c_{\alpha},
\eeqa
where $\cal{R}$ is the conversion factor to represent 
the response of the distortion of the galaxy image to 
a small shear \cite{Bernstein:2001nz},
$\gamma_{{\rm true}, \alpha}$ is the true value of cosmic shear,
and $m$ and $c_{\alpha}$ are
the multiplicative and additive biases to assess possible 
systematic uncertainty in galaxy shape measurements. In practice,
before employing the conversion in Eq.~(\ref{eq:gamma2kappa}),
one must first construct the smoothed shear field 
on grids \cite{Seitz:1994gz},
\beqa
\gamma_{{\rm sm}, \alpha}(\bd{\theta}) 
= \frac{\sum_{i} w_{i} 
\left(\tilde{\gamma}_{i, \alpha}-c_{i, \alpha}\right)\,  
W(\bd{\theta}-\bd{\theta}_{i})}
{\sum_{i} w_{i} (1+m_{i}) \, W(\bd{\theta}-\bd{\theta}_{i})},
\label{eq:smoothed_shear}
\eeqa
where $\bd{\theta}_{i}$ locates the position of the $i$-th 
source galaxy, $w_{i}$ represents the inverse variance weight 
for the $i$-th source galaxy, and $W(\bd{\theta})$ is 
assumed to be a Gaussian
\beqa
W(\bd{\theta}) = \frac{1}{\pi\theta_{G}^2}\exp\left(-\frac{\theta^2}{\theta_{G}^2}\right). \label{eq:filter_for_shear}
\eeqa
Using Eqs.~(\ref{eq:gamma2kappa}) and (\ref{eq:smoothed_shear}),
one can derive the smoothed convergence field from the observed imaging 
data through a Fast Fourier Transform (FFT).

The resulting field is mathematically 
equivalent to the filtered convergence field \cite{Schneider:1996ug}
which is defined as
\beqa
\kappa_{\rm sm}(\bd{\theta})
= \int {\rm d}^2\phi\, \kappa(\bd{\phi})U_{\kappa}(\bd{\theta}-\bd{\phi}),
\label{eq:smoothed_kappa}
\eeqa
where the smoothing filter for $\kappa$ 
can be expressed as 
\beqa
U_{\kappa}(\theta) =2\int_{\theta}^{\infty}{\rm d}\theta^{\prime}\,\frac{W(\theta^{\prime})}{\theta^{\prime}}-W(\theta).
\label{eq:filter_for_kappa}
\eeqa


\section{\label{sec:data}DATA}

\begin{figure*}
\begin{center}
       \includegraphics[clip, width=2.1\columnwidth, bb=0 0 1024 768]
       {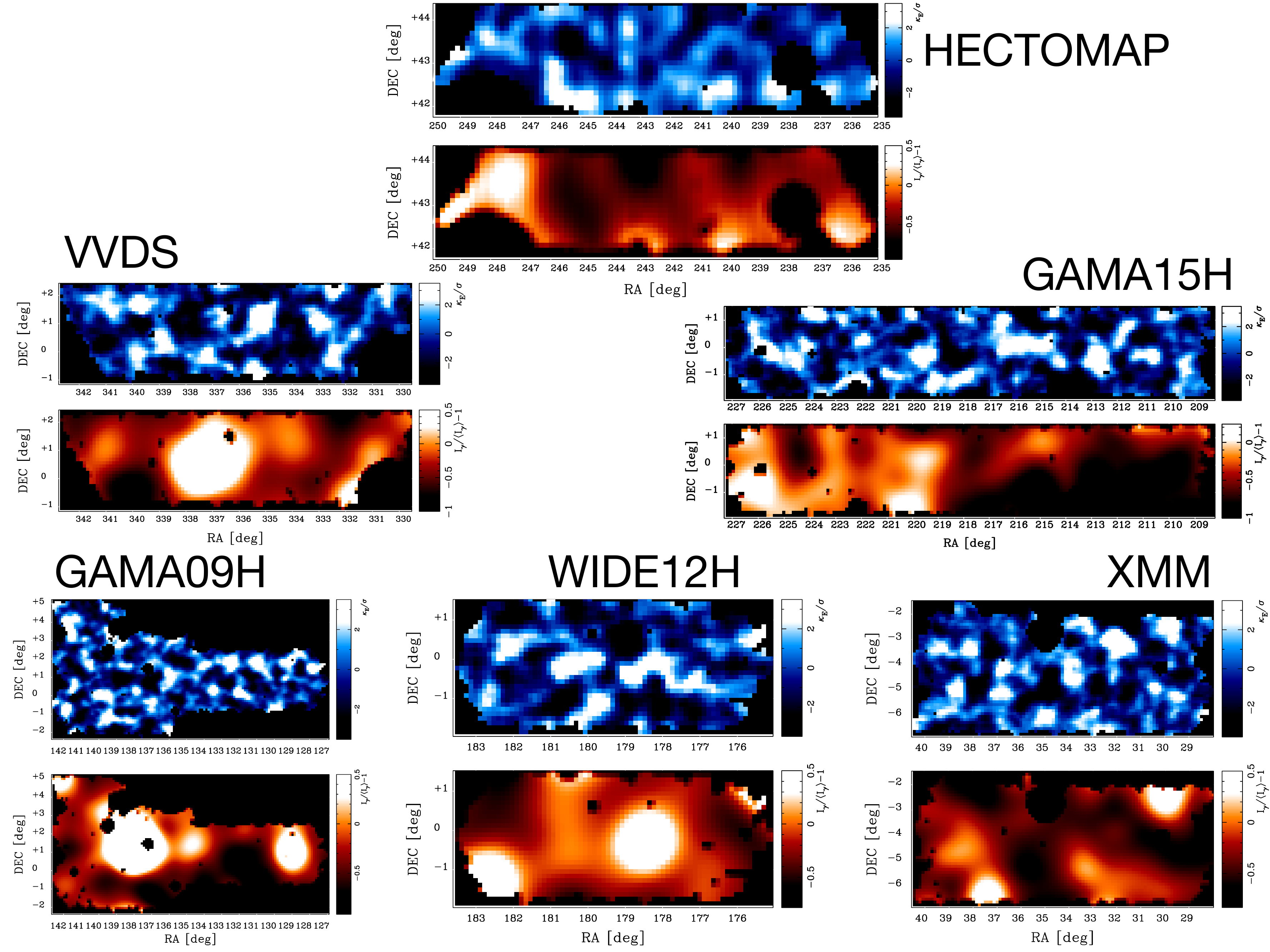}
     \caption{
     \label{fig:map}
     We plot the smoothed convergence field (upper panels) and 
     extragalactic gamma-ray background intensity (EGB; 
     bottom panels) for 6 HSC patches as labeled.
     In this figure, we set the smoothing scale to be 
     $\theta_G = 40$ arcmin in Eq.~(\ref{eq:filter_for_shear})
     and use all source galaxies 
     (no photometric redshift selections are applied).
     We also apply a Gaussian smoothing of 40 arcmin for 
     the EGB maps. Brighter regions correspond to  
     denser in mass or higher in 
     $\gamma$-ray intensity, whereas darker regions are 
     for more under-dense in mass or fainter in $\gamma$-ray
     intensity.
     }
    \end{center}
\end{figure*}

\subsection{\emph{Fermi}-LAT}\label{subsection:FermiDataSelection}
We use $\sim 7$ years (from August 4, 2008 to September 4, 2015) of Pass~8 {\tt ULTRACLEANVETO} photons with reconstructed energy in the 1~--~500 GeV range. Photons detected at zenith angles larger than 90$^\circ$ were excised in order to limit the contamination from $\gamma$-rays generated by cosmic-ray interactions in the Earth limb. Moreover, data were filtered removing time periods when the instrument was not in sky-survey mode. \emph{Fermi} Science Tools v10r0p5 and instrument response functions (IRFs) {\tt P8R2\_ULTRACLEANVETO\_V6} were used for this analysis. We included photons within a $20^{\circ} \times 20^{\circ}$ square region encompassing each of the HSC fields of view. Figure~\ref{fig:map} shows smoothed EGB maps observed by the LAT above 1 GeV (details of the \emph{Fermi}-LAT analysis and data processing are provided in the appendix).

\subsection{Subaru Hyper-Suprime Cam}
\label{subsec:hsc}

Hyper Suprime-Cam (HSC) is a wide-field imaging camera 
on the prime focus of the 8.2m Subaru telescope \cite{Aihara:2017paw,
2018PASJ...70S...1M,
2018PASJ...70S...2K,
2018PASJ...70S...3F}
(also see Kawanomoto,~S.,~et~al.~2018 in prep.). 
Among three layers in the HSC survey, 
the Wide layer will cover 1400 ${\rm deg}^2$ 
in five broad photometric bands (grizy) over 5--6 years, 
with excellent image quality of sub-arcsec seeing.
In this paper, we use a catalog of galaxy shapes 
that has been generated for cosmological weak 
lensing analysis in the first year data release. 
We denote the data as HSCS16A. The details of galaxy 
shape measurements and catalog information are found 
in Ref.~\cite{Mandelbaum:2017dvy}.

In brief, the HSCS16A shape catalog contains the shapes 
of $\sim$12 million galaxies selected from 137 ${\rm deg}^2$ 
measured with the re-Gaussianization method \cite{Hirata:2003cv}.
The shape measurement is performed in the coadded 
$i$-band image and calibrated by simulated galaxy images 
similar to those used in {\tt GREAT3} \cite{Mandelbaum:2014fta}. 
The image simulation enables to carry out reliable shear 
calibration, since it takes into account realistic HSC PSFs 
and reproduces the observed distribution of galaxy properties 
with remarkable accuracy \cite{Mandelbaum:2017ctf}.
We apply a conservative galaxy selection criteria for the first 
year science, e.g., $S/N\ge10$ and $i\le24.5$, giving an average 
raw number density of galaxies $\bar{n}\sim25\, {\rm arcmin}^{-2}$.
The HSCS16A dataset consists of 6 patches;
XMM, GAMA09H, GAMA15H, HECTOMAP, VVDS, and WIDE12H. 
While we present covergence maps for these individual
patches separately, 
we combine our results on cross-correlations for all 6 patches.
Accurate photometric redshifts in the HSCS16A are suitable
for tomographic approach as in Refs.~\cite{Camera:2014rja, Troster:2016sgf}.
While photometric redshifts are measured for the HSC galaxies 
using several different techniques, throughout this paper we 
use the {\tt MLZ} photometric redshifts \cite{Tanaka:2017lit}.
Individual HSC galaxies have a posterior probability distribution 
function (PDF) of redshift estimated by {\tt MLZ}.
We define the point estimate of each galaxy's redshift 
where its PDF takes the maximum.
We use the PDF information to evaluate the mean redshift 
as $\langle z \rangle = \int {\rm d}z\, z \sum_{i}^{N_{\rm gal}}\, P_i(z) /N_{\rm gal}$,
where $N_{\rm gal}$ is the number of source galaxies and 
$P_i(z)$ represents the PDF of photometric redshift for $i$-th galaxy.
When using all galaxies in the HSC shape catalog, 
we obtain the mean source redshift $z_{\rm m} = 0.96$.
For our lensing tomography analysis, 
we divide the source galaxies into three bins 
using their point estimates $z_{\rm photo}$ as
$0.3<z_{\rm photo}<0.8$, $0.8<z_{\rm photo}<1.2$,
and $1.2<z_{\rm photo}<1.9$,
and the corresponding mean source redshifts are 
$z_{\rm m}=0.60$, $1.00$, and $1.42$, respectively.
The true redshift distribution of the source galaxies is 
commonly approximated as the sum of the PDFs estimated from {\tt MLZ}.
We use the sum as the source redshift distribution 
in our theoretical model of weak lensing signals (see Eq.~[\ref{eq:lens_kernel}]).

We then reconstruct the smoothed convergence field from 
the HSCS16A data as described in Section~\ref{subsec:reconst_kappa}.
We follow a similar approach to that of Ref.~\cite{Oguri:2017vrv}.
Adopting a flat-sky approximation, 
we first create a pixelized shear map for 
each of the 6 patches on regular grids with 
a grid size of 0.1 deg. We then apply the FFT and perform 
convolution in Fourier space to obtain the smoothed convergence field, 
which is referred to as the E-mode convergence $\kappa_E$
in the literature. 
The imaginary part of the reconstructed convergence 
corresponds to the B-mode convergence, $\kappa_B$, 
which can be an indicator of the existence of certain types 
of residual systematics in our weak lensing measurements.

In actual observations, there are missing galaxy shear data 
due to bright star masks and edges.
Applying our method directly to such regions likely generates
noisy maps. We thus determine the mask regions for each convergence map
by using the smoothed number density map of the input galaxies
with the same smoothing kernel as in Eq.~(\ref{eq:filter_for_shear}). 
Then we mask all pixels with the smoothed
galaxy number density less than 0.5 times the mean number density.
In addition, we apply a conservative mask
of $|b|<30$ deg about the Galactic plane
in the cross-correlation analysis presented in
Section~\ref{subsec:est_cross_corr}.
Basic characteristics of the HSCS16A data are 
summarized in Table~\ref{tab:hsc_survey}.
Figure~\ref{fig:map} shows our reconstructed EGB
(bottom panels) and convergence fields (upper panels)
when we set the smoothing size of 
$\theta_G = 40$ arcmin.

\begin{table*}[!t]
\begin{center}
\scalebox{0.85}[0.85]{
\begin{tabular}{|c|r|r|r|r|r|}
\tableline
Name of patch & 
$n_{\rm gal}$ (No selection in $z_{\rm photo}$) &
$n_{\rm gal}$ ($0.3<z_{\rm photo}<0.8$) &
$n_{\rm gal}$ ($0.8<z_{\rm photo}<1.2$) &
$n_{\rm gal}$ ($1.2<z_{\rm photo}<1.9$) &
Effective area (${\rm deg}^2$)  \\ \tableline
GAMA09H & 16.8 & 7.93 & 3.84 & 2.72 & 21.53 \\
GAMA15H & 21.5 & 9.32 & 5.29 & 3.85 & 38.04 \\
HECTOMAP & 18.1 & 8.10 & 4.65 & 3.04 & 18.78 \\
VVDS & 19.2 & 8.30 & 4.71 & 3.56 & 24.49 \\
WIDE12H & 23.0 & 10.1 & 5.91 & 4.13 & 16.16 \\
XMM & 19.9 & 8.61 & 5.12 & 3.57 & 34.52 \\
\tableline
\end{tabular}
}
\caption{
\label{tab:hsc_survey}
Summary of lensing data used in this paper.
$n_{\rm gal}$ represents the weighted number density of source galaxies in unit of 
${\rm arcmin}^{-2}$, and the last column shows the effective survey area. 
The effective survey area is defined 
by the unmasked pixels in our EGB and convergence maps.
The definition of mask in our analysis is found in Section~\ref{subsec:hsc}.
}
\end{center}
\end{table*}

\subsection{Estimator of cross correlation}
\label{subsec:est_cross_corr}

In order to measure the $\kappa - \gamma$ cross-correlation, 
we introduce the following simple statistic
for our smoothed maps,
\beqa
\langle \kappa_{E} I_{\gamma} \rangle (\theta_{G})
\equiv \frac{\sum_{i} \kappa_{\rm sm}(\bd{\theta}_{i}) I_{\gamma,{\rm sm}} (\bd{\theta}_{i})}{\sum_{i}1},
\label{eq:cross_corr_est}
\eeqa
with
\beqa
I_{\gamma, {\rm sm}}(\bd{\theta})
= \int {\rm d}^2\phi\, I_{\gamma}(\bd{\phi})W(\bd{\theta}-\bd{\phi}).
\label{eq:smoothed_Igamma}
\eeqa
Here, $\kappa_{\rm sm}(\bd{\theta}_{i})$
represents the smoothed convergence map
reconstructed as in Section~\ref{subsec:reconst_kappa} 
(or defined in Eq.~[\ref{eq:smoothed_kappa}]),
and the sum in Eq.~(\ref{eq:cross_corr_est})
is taken over all unmasked pixels for individual HSCS16A fields.
Hence, the quantity $\langle \kappa_{E} I_{\gamma} \rangle$ is a function of
the smoothing scale $\theta_G$. It also depends on the photometric-redshift selection of source galaxies in gravitational lensing analysis.
We also examine the similar quantity of 
$\langle \kappa_{B} I_{\gamma} \rangle$
to study residual systematic effects in
the galaxy shape measurement.

To estimate the statistical error, we utilize 200 mock shear catalogs for the HSCS16A\footnote{
We chose the number of mock catalogs so that the inverse covariance matrix in our analysis has a $\sim10\%$ accuracy.
Ref.~\cite{Hartlap:2006kj} examine the bias in estimating 
the inverse covariance matrix as a function of independent 
random realizations. They find that the amplitude of bias depends 
on the ratio of the number of bins (data vector variables)
to the number of data sets.
There are 16 data variables 
in our cross correlation analysis.
With 200 random realizations, we expect that the bias in the inverse covariance matrix is  
as small as $200/(200-16-1)\simeq1.09$.
}. 
We follow the method developed in 
Refs.~\cite{Oguri:2017vrv, Shirasaki:2013zpa}
to create realistic mock catalogs that
incorporate the features of actual data
and ray-tracing simulations 
(also see Ref.~\cite{Mandelbaum:2017ctf} and Shirasaki~et~al.~in prep).
We use the all-sky ray-tracing simulations of gravitational lensing in Ref.~\cite{Takahashi:2017hjr}.
Each mock data consists of 38 different source planes each
separated by a comoving separation of $\Delta \chi=150\, h^{-1}{\rm Mpc}$, 
covering source planes up to redshift of 5.3. 
The angular resolution 
is set to be 0.43 arcmin. 
We use the observed photometric 
redshifts and angular positions of real galaxies as in Ref.~\cite{Oguri:2017vrv}. 
In short, we perform the following procedures: (1) we assign each
real HSC galaxy to the nearest angular pixel in the nearest redshift source plane, 
(2) we randomly rotate the orientation of the galaxy to remove the real lensing effect,
(3) we simulate the lensing distortion effect at the source position 
by adding the local lensing shear and the intrinsic shape, 
(4) we include the additional variance due to measurement error,
and 
(5) repeat the procedures (1)-(4) for all the source galaxies.
Note that, when generating a realization, we randomly sample the source redshift from the posterior PDF of photometric redshift for each galaxy. 
Our mock catalogs include directly the properties of source galaxies 
(e.g., magnitudes, ellipticities and spatial variations in the number densities), 
statistical uncertainties in photometric redshifts, and also the survey geometry.
We use 10 full-sky simulations to extract 200 mock HSCS16A catalogs.
 
For each patch of the HSCS16A data, 
we evaluate the cross-correlation Eq.~(\ref{eq:cross_corr_est})
and 
estimate their errors by using
the standard deviation from our 200 mock catalogs.
Using the measured correlations and the standard deviations for a total of 6 patches, 
we combine them by the inverse-variance weighted method \cite{Shirasaki:2016kol,Oguri:2017vrv}.
The same procedures are also applied to the 200 mock catalogs, 
in order to estimate the covariance of 
$\langle \kappa_{E}I_{\gamma} \rangle$ in the HSCS16A 
among different $\theta_{G}$
and source redshift selections.
To justify our procedure of combining the cross correlations, 
we have checked the field variation in EGB mean intensity above 1 GeV
and that of the mean redshift of HSC galaxies.
We find $20-30\%$ differences in the EGB mean intensity
and less than $5\%$ difference in the mean redshift among different patches.
These differences are small enough compared to the current statistical uncertainty in the cross correlation measurement.

\section{\label{sec:model}ANALYTIC MODEL}

\subsection{Formulation}
\label{subsec:form}
Here we derive the expectation value of 
Eq.~(\ref{eq:cross_corr_est})
in the standard structure formation model.
Let us begin with the two-point cross-correlation function between 
$\kappa_{\rm sm}$ and $I_{\gamma, {\rm sm}}$ in real space.
Under a flat-sky approximation, it is given by
\beqa
\langle \kappa_{\rm sm}(\bd{\theta}_{1}) I_{\gamma, {\rm sm}} (\bd{\theta}_{2})\rangle
&=&
\int\frac{{\rm d}^2\ell_1}{(2\pi)^2} \int\frac{{\rm d}^2\ell_2}{(2\pi)^2}\,
e^{i\bd{\ell}_{1}\cdot\bd{\theta}_{1}-i\bd{\ell}_{2}\cdot\bd{\theta}_{2}}
\nonumber \\
&&
\,\,\,\,\,
\,\,\,\,\,
\,\,\,\,\,
\,\,\,\,\,
\times
\langle \hat{\kappa}_{\rm sm}(\bd{\ell}_{1}) \hat{I}^{*}_{\gamma, {\rm sm}} (\bd{\ell}_{2})\rangle, 
\nonumber \\
&=&\int \frac{{\rm d}^2\ell_1}{(2\pi)^2}\, 
e^{i\bd{\ell}_{1}\cdot\bd{\theta}_{12}}C^{\kappa I_{\gamma}}(\ell_1)
\nonumber \\
&&
\,\,\,\,\,
\,\,\,\,\,
\,\,\,\,\,
\times
\hat{W}(\ell_1, \theta_G)\hat{U}_{\kappa}(\ell_1, \theta_G), \label{eq:cross_corr}
\eeqa
where characters with hats represent quantities in Fourier space,
$\delta^{(2)}(\bd{x})$ is the dirac delta function in two-dimensional space,
$\bd{\theta}_{12} = \bd{\theta}_{1} - \bd{\theta}_{2}$,
and we define the cross power spectrum $C^{\kappa I_{\gamma}}$ as,
\beqa
\langle \hat{\kappa}(\bd{\ell}_{1}) \hat{I}_{\gamma} (\bd{\ell}_{2})\rangle 
&\equiv& C^{\kappa I_{\gamma}}(\ell_1) (2\pi)^2 \delta^{(2)}(\bd{\ell}_1-\bd{\ell}_{2}).
\label{eq:cross_power}
\eeqa
Here we use the relations of Eqs.~(\ref{eq:smoothed_kappa}) 
and (\ref{eq:smoothed_Igamma}) in Fourier space,
\beqa
\hat{\kappa}_{\rm sm}(\bd{\ell}) = \hat{U}_{\kappa}(\ell, \theta_G) \hat{\kappa}(\bd{\ell}), \\
\hat{I}_{\gamma,{\rm sm}}(\bd{\ell}) = \hat{W}(\ell, \theta_G) \hat{I}_{\gamma}(\bd{\ell}),
\eeqa
where $W$ and $U_{\kappa}$ are defined as Eqs.~(\ref{eq:filter_for_shear}) and (\ref{eq:filter_for_kappa}), respectively.
For a given filter of $f(\theta)$, its Fourier counterpart is written as,
\beqa
\hat{f}(\ell) = \int_{0}^{\infty} 2\pi\theta {\rm d}\theta\, f(\theta)\, J_{0}(\ell\theta),
\eeqa
where $J_{0}(x)$ is the zero-th Bessel function.
Hence, 
we can obtain the expectation value of Eq.~(\ref{eq:cross_corr_est})
by setting $\bd{\theta}_{12}=0$ in the right-hand side of Eq.~(\ref{eq:cross_corr}).

The cross power spectrum $C^{\kappa I_{\gamma}}$ can be expressed as \cite{Fornengo:2013rga, Camera:2012cj, Camera:2014rja},
\beqa
C^{\kappa I_\gamma}(\ell)
&=& \sum_{X} \int \frac{{\rm d}\chi}{r^2(\chi)}
W_{g,X}(\chi) W_{\kappa}(\chi) \nonumber \\
&&
\,\,\,\,\,
\,\,\,\,\,
\,\,\,\,\,
\,\,\,\,\,
\times
P_{mg,X}\left(\frac{\ell+1/2}{r(\chi)}, z(\chi)\right), \label{eq:analytical_cross_power}
\eeqa
where 
$P_{mg,X}$ represents the three-dimensional cross power spectrum
between the matter density and the number density of astrophysical source population 
$X$ in the EGB. Here we ignore possible contributions from intrinsically diffuse processes such as dark-matter annihilation and decay\footnote{
Cosmological constraints of dark-matter annihilation with our measurements
are provided in the Appendix, while our method seems more suitable for studying astrophysical contributions in the EGB.
}.
We also note that the effect of the energy dependent $\gamma$-ray PSF is
properly included as in Ref.~\cite{Shirasaki:2014noa},
when we compare our model with the observed cross correlations.

\begin{figure}
\begin{center}
       \includegraphics[clip, width=0.8\columnwidth, bb= 0 0 510 461]
       {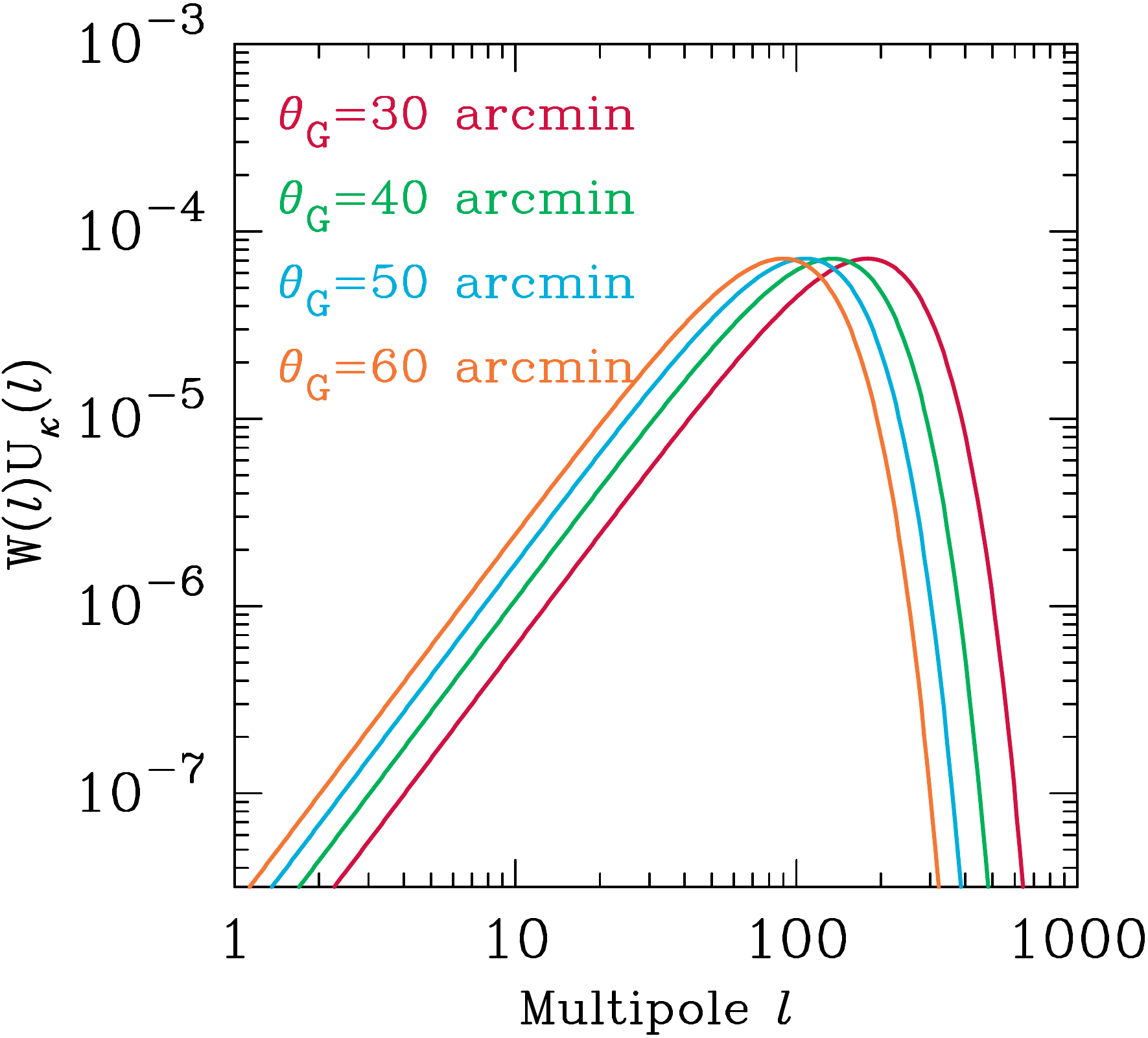}
     \caption{
     \label{fig:kernel_f}     
     Effective smoothing filter in our cross-correlation analysis.
     The horizontal axis shows the product of two filters of interest in Fourier space.
     The difference in colors represent the case 
     with different smoothing scales $\theta_{G}$.
     Over the range of $\theta_{G}\sim30$ to 60 arcmin, our filter effectively
     extracts information of $\ell\sim100$, i.e., clustering in degree scales.
  }
  \end{center}
\end{figure}

\begin{figure}
\begin{center}
       \includegraphics[clip, width=0.8\columnwidth, bb= 0 0 496 503]
       {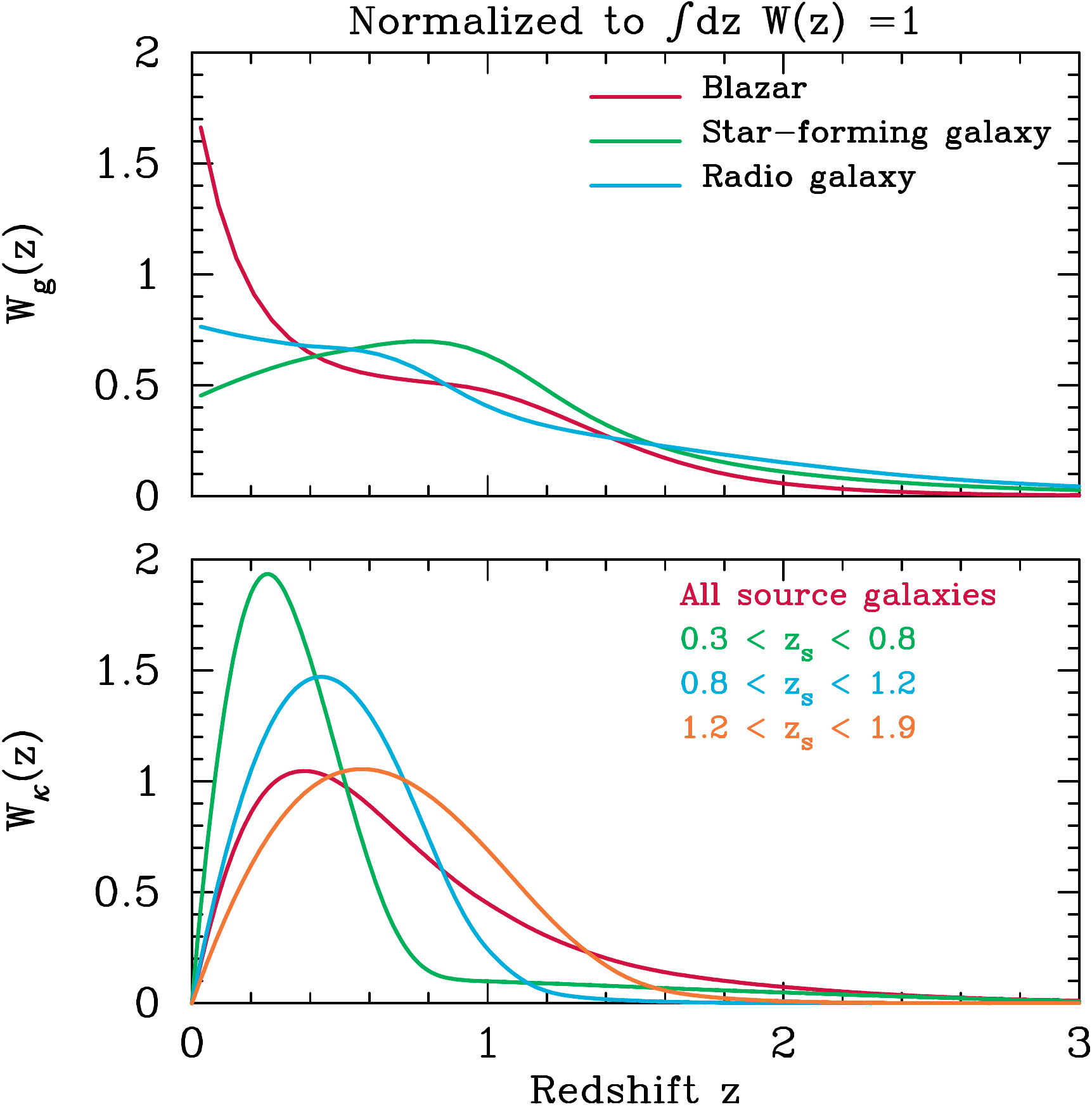}
     \caption{
     \label{fig:windows}
     Window function in redshift direction.
     Upper panel represents the window for EGB
     and different colored lines show the window for different astrophysical sources.
     We work with the $\gamma$-ray energy range of 1--500 GeV.
     The bottom panel is the window for lensing convergence.
     Here, we consider the four cases with different 
     source-galaxy redshift selections in the HSCS16A data:  
     no selections (red), $0.3<z_{\rm photo}<0.8$ (green),
     $0.8<z_{\rm photo}<1.2$ (cyan), and 
     $1.2<z_{\rm photo}<1.9$ (orange).
     Note that every window function is normalized to $\int {\rm d}z\, W(z)=1$.
  } 
    \end{center}
\end{figure}

Before going to the detailed computation,
we describe our smoothing filters 
in Fourier space and the window function along a line of sight.
Figure~\ref{fig:kernel_f} shows the effective smoothing filter 
$\hat{W}\hat{U}_{\kappa}$ in Eq.~(\ref{eq:cross_corr}).
According to this figure, our filters in Eqs.~(\ref{eq:filter_for_shear}) and (\ref{eq:filter_for_kappa}) can be regarded as high-pass and low-pass in smoothing.
In the range of $\theta_G=30-60$ arcmin, we can efficiently extract the information
of clustering signature between Fourier modes with narrow range of $\ell\sim100$.
Furthermore, Figure~\ref{fig:windows} summarizes the relevant 
window functions along a line of sight in our analysis.
Because of the nature of gravitational lensing,
we can effectively remove the contribution from very low-redshift structures.
The effective window function in the computation of $C^{\kappa I_\gamma}$
is broad in redshift, but we can find that the effective redshift
will be of the order of $\sim0.5$ corresponding to $\chi\sim1$ Gpc.
Therefore, the main contribution in Eq.~(\ref{eq:analytical_cross_power})
will come from the clustering between matter density and $\gamma$-ray sources
at $\sim10$ Mpc, 
where the linear theory in structure formation is valid with reasonable accuracy.

\subsection{A model of bias of astrophysical sources}
\label{subsec:fiducial_model}

We introduce our fiducial model of astrophysical sources.
Our model is largely based on the one developed in Ref.~\cite{Camera:2014rja}.
This model is found to be consistent with the observed cross correlation signals between the IGRB 
and the gravitational-lensing effect in the CMB \cite{Fornengo:2014cya}.
As seen above, the expected correlation can be mostly 
set by the linear part of the clustering. 
In this case, the three-dimensional cross power spectrum
can be approximated as 
$P_{mg,X}(k, z) = b_{{\rm eff}, X}(z)P_{L}(k,z)$,
where $P_{L}(k,z)$ is the linear matter power spectrum at redshift $z$
and $b_{{\rm eff},X}$ is the effective bias for source population $X$.
We follow the standard halo-model approach to evaluate $b_{{\rm eff},X}$ \cite{Camera:2012cj, Camera:2014rja}.
Assuming correlation between the $\gamma$-ray luminosity
and host halo mass $M_{h}$ of $\gamma$-ray source, one can find,
\beqa
b_{{\rm eff},X}(z) = 
\frac{\int {\rm d}L_{\gamma}\, b_{h}(M_{h, X}(L_{\gamma},z),z) 
L_{\gamma}\Phi_{X}(L_{\gamma},z)}{\int {\rm d}L_{\gamma}\, L_{\gamma}\Phi_{X}(L_{\gamma},z)},
\eeqa
where $\Phi_{X}$ is the luminosity function of the source population $X$ (here we already perform the integral over the photon index $\Gamma_X$ if needed), 
$M_{h, X}(L_{\gamma},z)$ represents the relation of host halo and the $\gamma$-ray luminosity, and $b_{h}$ is the linear halo bias.
In this paper, we adopt the model of $b_{h}$ in Ref.~\cite{Sheth:1999mn}.

\begin{figure}
\begin{center}
       \includegraphics[clip, width=0.9\columnwidth, bb=0 0 525 507]
       {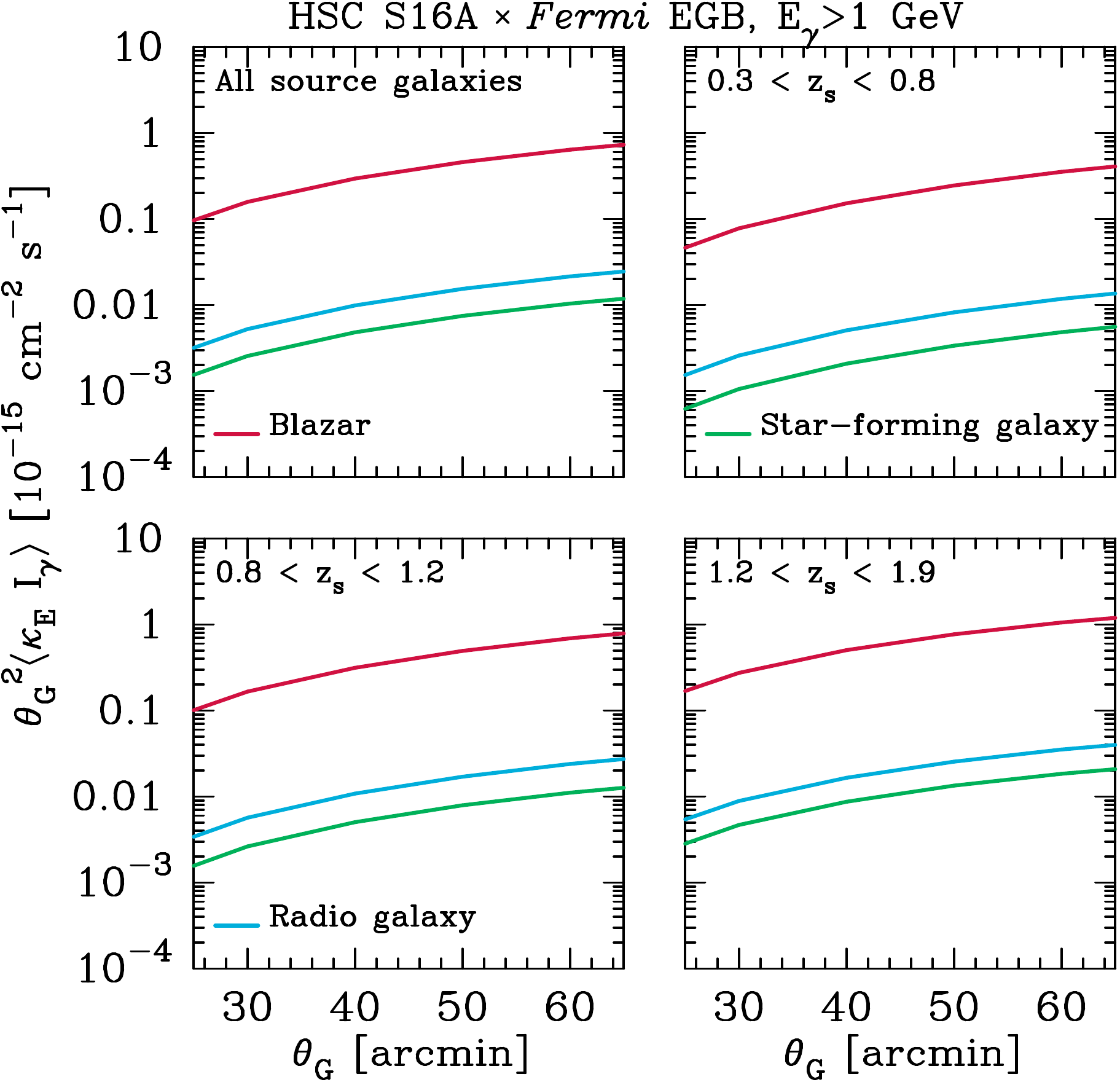}
     \caption{
     \label{fig:expected_signal}
     Expected cross correlation signal between the EGB
     and lensing convergence.
     Each panel shows the results using different source galaxies 
     selected by their photometric redshift in the lensing analysis.
     In each panel, the red line represents the contribution from correlation to 
     blazars, the green is for star-forming galaxy
     and the cyan is for radio galaxy.
     The blazar contribution is expected to be dominant,
     indicating that we can constrain the physical property of blazars
     with our cross correlation analysis.
  } 
    \end{center}
\end{figure}

For blazars, we use the model in Ref.~\cite{Hutsi:2013hwa}, 
which has introduced 
a simple power-law scaling 
$M_h=A(z){\cal L}_{\rm X-ray}^{B(z)}$,
to the X-ray luminosity ${\cal L}_{\rm X-ray}$
so that the model can reproduce well 
the abundance of X-ray selected AGNs 
at different redshifts
(see their Fig.~6 and Table 1).
To relate the X-ray luminosity to the $\gamma$-ray luminosity,
we introduce the relation between
the bolometric blazar luminosity $P$ 
and disk X-ray luminosity as $P=10^q {\cal L}_{\rm X-ray}$ \cite{Inoue:2008pk}.
The parameter $q$ is set to be 4.21 as 
follows in Ref.~\cite{Harding:2012gk}.

For star-forming and radio galaxies, constructing the possible relation between $M_h$ and $L_{\gamma}$ is more uncertain 
since there exists limited data.
In the case of star-forming galaxy, we adopt 
the benchmark model in Refs.~\cite{Camera:2012cj,Camera:2014rja}
$M_{h} = 10^{12}\, M_{\odot} (L_{\gamma}/10^{39}\, {\rm erg}\, {\rm s}^{-1})^{1/2}$,
where it is evaluated with reasonable approximation
that the mass associated to the maximum luminosity will not exceed a maximum galactic mass.
Ref.~\cite{Camera:2014rja} also constructed a model for the $M_{h}-L_{\gamma}$ relation for radio galaxies by using data 
from 12 available AGNs 
and possible relation in mass between dark-matter halo and 
super-massive black hole (SMBH) \cite{Hutsi:2013hwa, Bandara:2009sd}.
The best-fit relation can be approximated 
as $M_{h} = 3.2\times 10^{12}\, M_{\odot} (L_{\gamma}/10^{41}\, {\rm erg}\, {\rm s}^{-1})^{0.11}$.
Although one can construct more physical models, such as 
examined in Ref.~\cite{Camera:2014rja}, we expect that 
our main results will be 
weakly sensitive to the choice of the $M_h-L_{\gamma}$ relation 
for star-forming and radio galaxies. 
This is because the dominant contribution to the EGB is expected 
to be from blazars, and the window function for blazars is much larger than that of other contributions at redshifts relevant in our analysis. Note that the window function in EGB $W_{g,X}$ is now constrained by the measurement of the EGB energy spectrum \cite{Ackermann:2014usa, Ajello:2015mfa}
and it seems difficult to increase the contribution from star-forming and radio galaxies furthermore in the range of 1-500 GeV. 

Figure~\ref{fig:expected_signal} shows the expected signal from
our fiducial model of astrophysical $\gamma$-ray sources.
The figure clearly demonstrates that the blazar population gives a much larger contribution to $\langle \kappa_E I_{\gamma} \rangle$ 
than the others. The difference can be a factor of $10-100$ in our lensing analysis. 
These results strongly motivate us to constrain the 
physical properties of blazars with our cross correlation analysis.
We emphasize that the expected signal is largely contributed by 
the structures at $z\sim0.5$,
because the cross correlation is proportional to the factor of $W_{g,X}\times W_{\kappa}$ (also see Figure~\ref{fig:windows} and Eq.~[\ref{eq:analytical_cross_power}]). Hence, the correlation can be used to probe the clustering of blazars at relatively high redshift, which is inaccessible 
through auto-correlation of the resolved blazers.

\begin{figure}
\begin{center}
       \includegraphics[clip, width=0.9\columnwidth, bb=0 0 524 507]
       {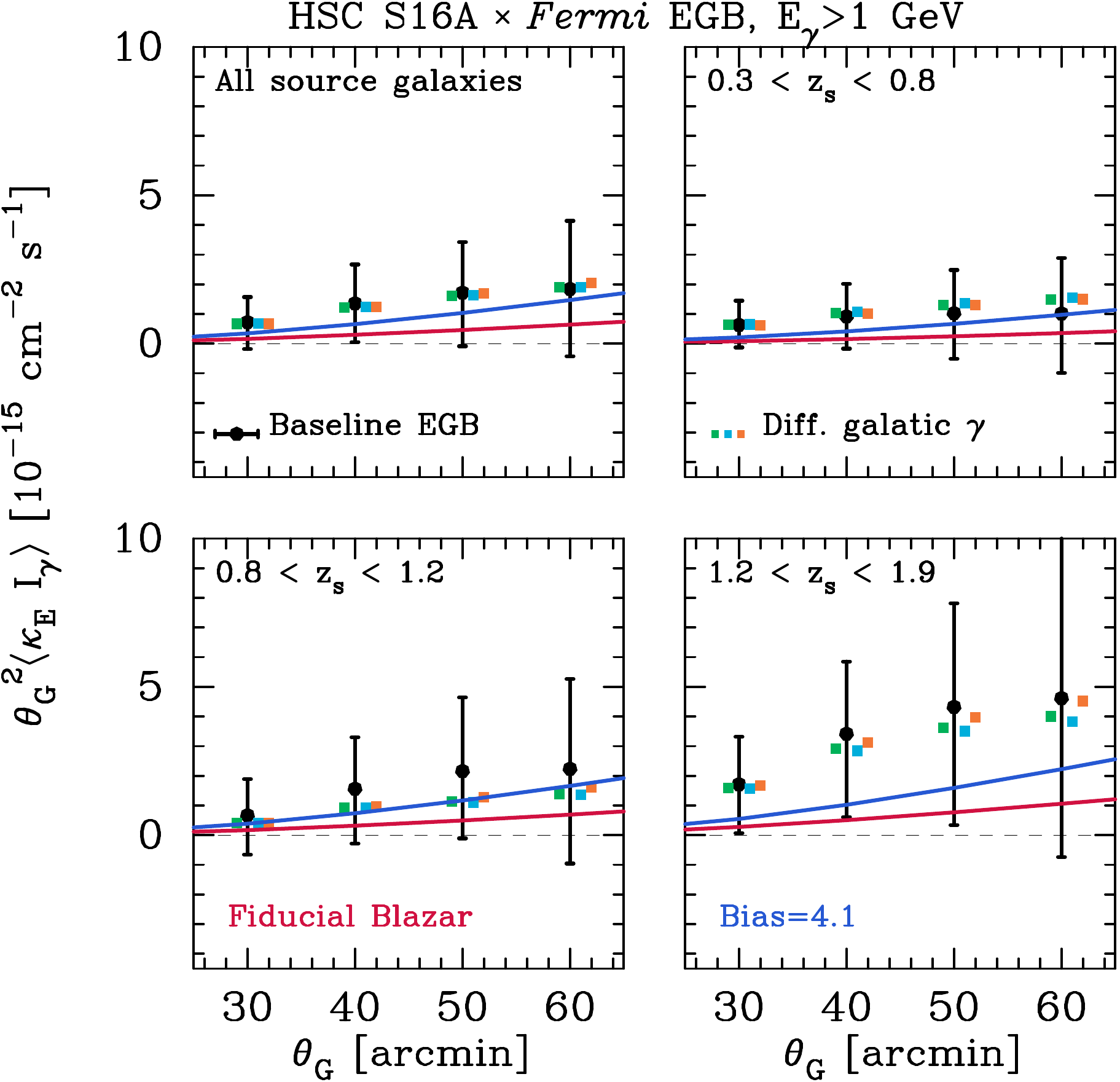}
       \includegraphics[clip, width=0.9\columnwidth, bb=0 0 524 507]
       {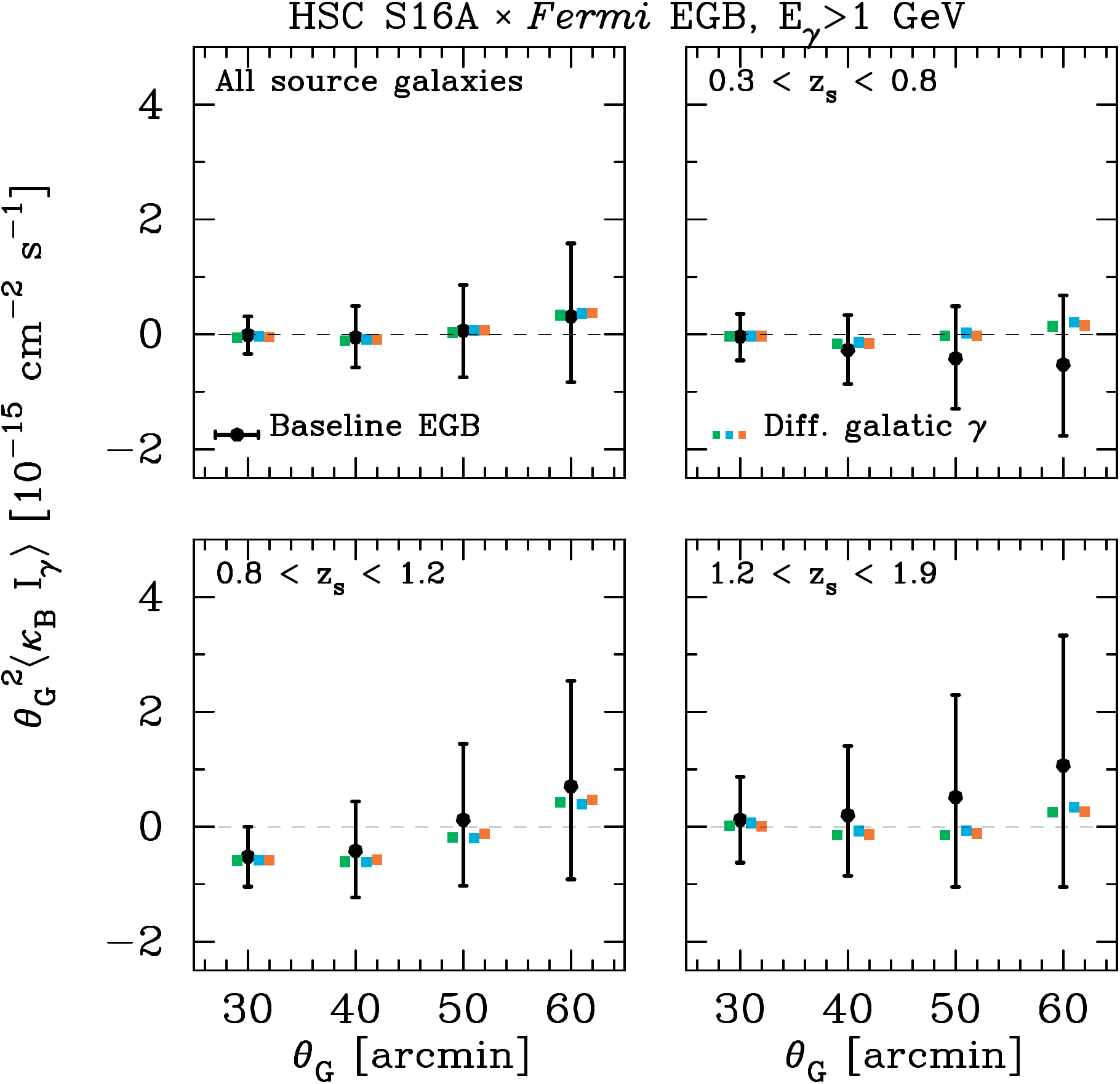}
     \caption{
     We plot the cross-correlations between
     Fermi-LAT and HSCS16A data.
     {\it Top}: The lensing convergence $\kappa_{\rm E}$ - EGB correlation as defined in 
     Eq.~(\ref{eq:cross_corr_est}).
     The four panels show the results obtained from different source-galaxy selections 
     by source redshift $z_{\rm photo}$: 
     all source galaxies (top left), 
     $0.3<z_{\rm photo}<0.8$ (top right),
     $0.8<z_{\rm photo}<1.2$ (bottom left), and
     $1.2<z_{\rm photo}<1.9$ (bottom right).
     The black points with error bars show the measurement using
     our baseline EGB maps, while the colored points with small offsets represent 
     the cases with different 
     galactic $\gamma$-ray templates used to construct the EGB 
     (see appendix \ref{sec:gammaanalysis} for details).
     Red lines correspond to our fiducial model of blazar clustering 
     (see Section~\ref{subsec:fiducial_model} for details).
     \change{Blue lines show the expected correlation signal with a constant blazar bias of 4.1.
     This large bias is found to be consistent with stacked analysis of $\kappa_{\rm E}$ 
     around resolved point sources (see Section~\ref{subsec:stack_kappa} for details).}
     {\it Bottom}: As above, but for the B-mode
     convergence $\kappa_B$, which is equivalent to the imaginary 
     part in Eq.~(\ref{eq:gamma2kappa}). 
     The small correlation between the EGB and $\kappa_B$
     suggests that the galaxy shape measurement is accurate enough.
     Note that each data point should be correlated with each other.
     \label{fig:cross_corr_EGB_HSCS16A}
  } 
    \end{center}
\end{figure}

\section{\label{sec:res}RESULT}

\subsection{Cross correlation}
\label{subsec:cross_corr}

Figure~\ref{fig:cross_corr_EGB_HSCS16A} summarizes 
our cross-correlation measurements. 
The error bars in this Figure are evaluated 
with 200 mock catalogs of galaxy shapes in HSCS16A.
The results of using different galactic $\gamma$-ray templates to construct the EGB (see appendix \ref{sec:gammaanalysis} for details) are shown by different colors. We confirm that the systematic uncertainty 
due to imperfect knowledge of galactic $\gamma$-ray components is smaller than the current statistical uncertainty.
This is because the ROI in the HSCS16A is sufficiently far from
the Galactic plane and the galactic $\gamma$-ray contribution
from $\pi^{0}$ decays can be well constrained.
Similar results have been found in Ref.~\cite{Shirasaki:2016kol}.
In addition, we also examine the cross correlation with 
the B-mode convergence $\kappa_B$ which is commonly used as 
an indicator of systematic uncertainties in the galaxy-shape measurement. 
The bottom set of 4 panels in Figure~\ref{fig:cross_corr_EGB_HSCS16A} show the
cross correlation with $\kappa_B$.
We define the significance of our measurements, i.e., null signals, as
\beqa
\chi_{0}^2 &=& \sum_{s, s^{\prime}}\sum_{i,j} 
{\bd C}^{-1}_{ij}(s,s^{\prime}) \nonumber \\
&&
\,\,\,\,
\,\,\,\,
\,\,\,\,
\,\,\,\,
\times
\langle\kappa_E I_\gamma \rangle (\theta_{G, i}; s)
\langle\kappa_E I_\gamma \rangle (\theta_{G, j}; s^{\prime}),
\label{eq:chi2}
\eeqa
where 
$\langle\kappa_E I_\gamma \rangle (\theta_{G, i}; s)$ represents
the cross correlation for smoothing scale $\theta_{G,i}$ 
with source galaxy selection $s$, 
and ${\bd C}(s,s^{\prime})$ is the covariance 
estimated from 200 mock catalogs.
We find that the cross correlation is consistent with 
null detection in all cases.
Table~\ref{tab:chi2} summarizes the $\chi_{0}^2$ in our measurements.

\begin{table}
\begin{center}
\begin{tabular}{|c|c|c|}
\tableline
Tomographic bins & 
$\langle \kappa_E I_\gamma \rangle$  &
$\langle \kappa_B I_\gamma \rangle$  \\ \hline \hline
No selection in $z_{\rm photo}$ 
& 1.84 (4) & 1.07 (4) \\
$0.3<z_{\rm photo}<0.8$ 
& 0.82 (4) & 1.14 (4) \\
$0.8<z_{\rm photo}<1.2$ 
& 1.96 (4) & 3.92 (4) \\
$1.2<z_{\rm photo}<1.9$ 
& 4.31 (4) &  0.61 (4) \\ \tableline
Combined 
& 7.41 (16) & 8.95 (16) \\ \tableline
\end{tabular}
\caption{
\label{tab:chi2} 
Summary of the significance of 
our cross-correlation measurements.
Second and third columns represent the $\chi_{0}^2$ defined in 
Eq.~(\ref{eq:chi2}) and the numbers in brackets show the degree of freedom in the analysis. 
Note that we take into account the off-diagonal terms in the covariance
between different smoothing scales and source-galaxy redshift selections.
}
\end{center}
\end{table}

\subsection{Implications for blazar-halo connection}
\label{subsec:implication}

\change{
Since weak gravitational lensing provides a direct, physical probe
of the cosmic matter density distribution,
our cross-correlation measurement can be used to determine or 
constrain the clustering bias of $\gamma$-ray sources
with respect to the underlying density field.
The top panels of figure~\ref{fig:cross_corr_EGB_HSCS16A}
compare the observational data
with our fiducial model predictions for the cross correlation 
expected from blazars (red lines). Interestingly, the data suggest even stronger
clustering than our fiducial model predicts. 
Our model of blazar clustering (Section~\ref{subsec:fiducial_model})
predicts a bias factor $b_{\rm eff}\simeq 2$ over the redshift range of interest,
which is consistent with the 
recent measurement of the blazar auto-correlation \cite{Allevato:2014qga}.
Although the statistical errors are still large, our cross-correlation analysis
suggests that the EGB sources may be a more biased tracer of the large-scale structure than expected.
}

\begin{figure}[t]
\begin{center}
        \includegraphics[clip, width=0.9\columnwidth]
       {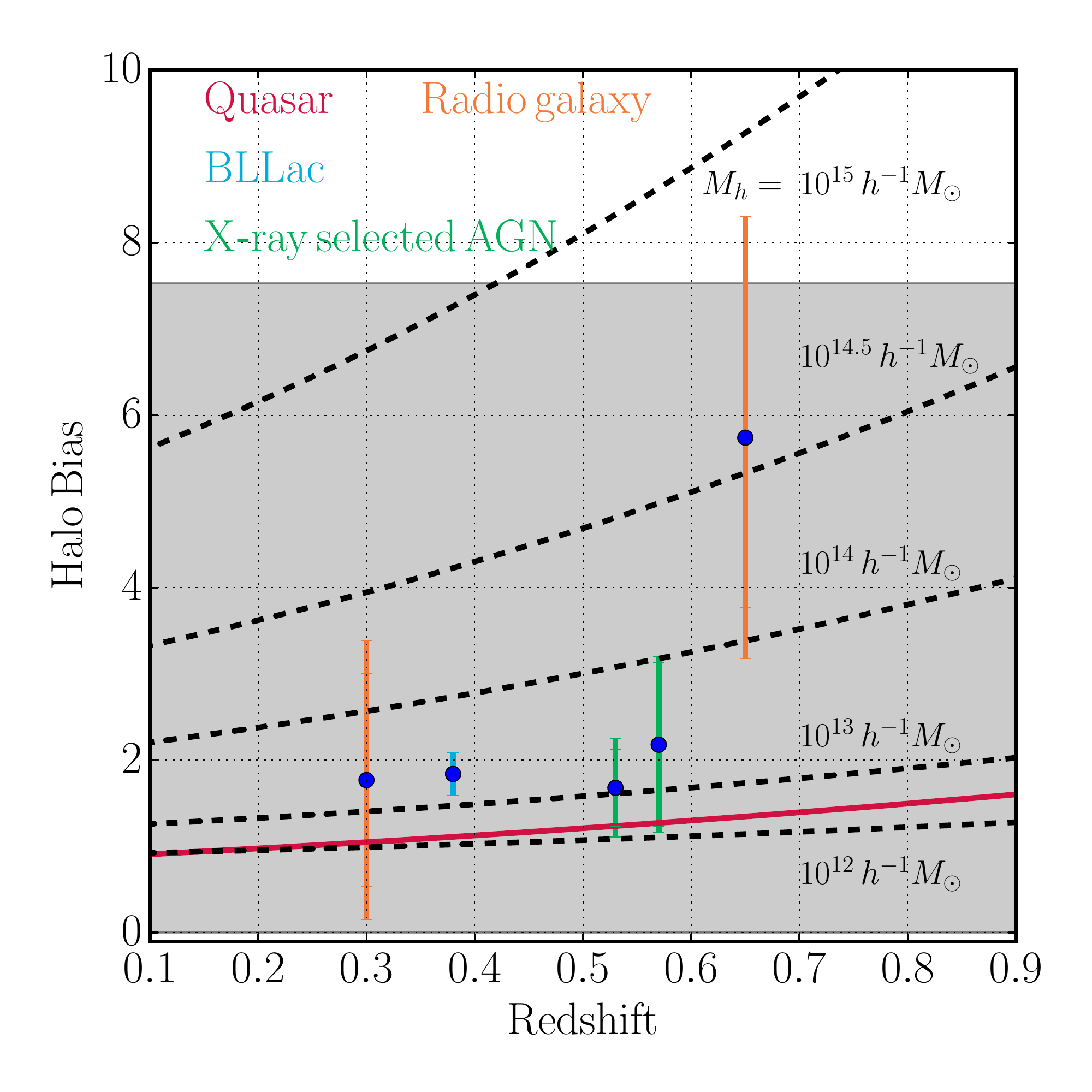}
     \caption{
     \label{fig:bias_constraint}
     Allowed blazar bias (gray shaded region) compared with 
     other clustering measurements of AGNs at different wavelengths.
     The red line shows the bias of quasars obtained from the 
     2dF QSO Redshift survey \cite{Croom:2004ui}, 
     the green data points are for the bias measurement of X-ray selected AGNs
     in the XMM Cosmic Evolution Survey \cite{2011ApJ...736...99A},
     the cyan data point represents the bias measurement of resolved blazars
     in the 2-year all-sky survey by Fermi-LAT
     \cite{Allevato:2014qga},
     and the orange data point is for the bias measurement of radio galaxies
     in the Very Large Array Faint Images of the Radio Sky at Twenty-cm survey (FIRST) \cite{Lindsay:2014rua}.
     For reference, the dashed line shows the linear halo bias with a fixed mass \cite{Sheth:1999mn}.
  } 
    \end{center}
\end{figure}

\change{
We can use the cross-correlation to place constraints on 
the blazar bias. Assuming the observed data follow multivariate Gaussian with 
covariance matrix $\bd C$, we introduce the $\chi^2$ statistics,
\beqa
\chi^2 &=& \sum_{s, s^{\prime}}\sum_{i,j} 
{\bd C}^{-1}_{ij}(s,s^{\prime}) 
\left\{\langle\kappa_E I_\gamma \rangle (\theta_{G, i}; s)
-\mu_{i,s}({\bd p)}\right\}
\nonumber \\
&&
\,\,\,\,
\,\,\,\,
\,\,\,\,
\,\,\,\,
\times
\left\{\langle\kappa_E I_\gamma \rangle (\theta_{G, j}; s^{\prime})
-\mu_{j,s^{\prime}}({\bd p)}\right\},
\label{eq:chi2_gen}
\eeqa
where $\mu_{i,s}$ represents theoretical template of interest
for observed $\langle\kappa_E I_\gamma \rangle (\theta_{G, i}; s)$.
In our theoretical model, we assume
a constant blazar bias over redshift, 
i.e., $b_{\rm eff}(z)=const$, for simplicity. 
The data consist of four bins in $\theta_{G}$ with four different
source-galaxy selections in the lensing analysis.
Hence, we have $4\times4=16$ degrees of freedom,
whereas there is only a single parameter in our model. 
The constraint on the parameter can be obtained 
by $\chi^2 -\chi_{0}^{2} < 1$, corresponding to a 
68\% confidence level. 
Using our measurements, we derive an upper limit 
on the blazar bias of $b_{\rm eff} < 7.53$.
We show the constraint in figure~\ref{fig:bias_constraint}
as the gray shaded region, 
and compare it with the bias measurements of AGNs selected at different wavelengths.
The overall uncertainty remains fairly large, both in the bias parameter and
in the source redshift, but it is intriguing that blazars contributing to the EGB
may likely reside in massive dark halos at $z \sim 0.3-0.7$.
}

\subsection{Stacked convergence profile around resolved $\gamma$-ray sources}
\label{subsec:stack_kappa}

\change{
As a cross-check of our findings, we also study the lensing signals
around \textit{resolved} $\gamma$-ray sources in the HSCS16A fields.
}

\subsubsection*{Estimator}
\change{
We stack the reconstructed convergence $\kappa_E$ from galaxy shapes in HSCS16A, and 
calculate the $\kappa_E$ profile around the resolved $\gamma$-ray sources.
The estimator of the stacked profile is given by
\beqa
\langle \kappa_E \rangle (\theta) &=& 
\frac{\sum_{i,j}\, w_{ij}\, n_{\gamma}(\bd{\phi}_{i})\, \kappa_{E}(\bd{\phi}_{j})}{\sum_{i,j}\, w_{ij} n_{\gamma}(\bd{\phi}_{i})} \nonumber \\
&&
\,\,\,\,\,\,\,\,
- \frac{\sum_{i,j}\, w_{ij}\, n_{\rm rand}(\bd{\phi}_{i})\, \kappa_{E}(\bd{\phi}_{j})}{\sum_{i,j}\, w_{ij} n_{\rm rand}(\bd{\phi}_{i})},
\label{eq:est_stack_kappa}
\eeqa
where $n_{\gamma}$ represents the number density field of $\gamma$-ray sources and 
$n_{\rm rand}$ is the number density field of random points.
In Eq.~(\ref{eq:est_stack_kappa}),
we set the linear angular binning with a bin width $\Delta\theta$
so as to be $w_{ij} = 1$ if $\theta-\Delta\theta/2 < |\bd{\phi}_{i}-\bd{\phi}_{j}|< \theta+\Delta\theta/2$
and $w_{ij} =0$ otherwise.
The stacked profile of B-mode convergence $\langle \kappa_B \rangle$ can be estimated similarly to Eq.~(\ref{eq:est_stack_kappa}).
Note that $\langle \kappa_B \rangle$ should vanish if there are no systematic effects in galaxy shape measurements.
}

\subsubsection*{Expectation value}
\change{
The expectation value of Eq.~(\ref{eq:est_stack_kappa}) is computed as
\beqa
\langle \kappa_E \rangle (\theta) = 
\int \frac{{\rm d}^2\ell}{(2\pi)^2}\, 
e^{i\bd{\ell}\cdot\bd{\theta}}C^{\kappa n_{\gamma}}(\ell)
\hat{U}_{\kappa}(\ell, \theta_G), \label{eq:cross_corr_stack}
\eeqa
where $\hat{U}_{\kappa}$ is the Fourier counterpart of Eq.~(\ref{eq:filter_for_kappa}).
We here define the cross power spectrum between lensing convergence and 
the photon number density fluctuation field as
\beqa
\langle \hat{\kappa}(\bd{\ell}_{1}) \hat{\delta}_{\gamma} (\bd{\ell}_{2})\rangle 
&\equiv& C^{\kappa n_{\gamma}}(\ell_1) (2\pi)^2 \delta^{(2)}(\bd{\ell}_1-\bd{\ell}_{2}),
\label{eq:cross_power_stack}
\eeqa
where $\delta_{\gamma} = n_{\gamma}/\bar{n}_{\gamma}-1$ and 
$\bar{n}_{\gamma}$ is the average angular number density of $\gamma$-ray sources.
Under the linear approximation, we express the cross power spectra as
\beqa
C^{\kappa n_{\gamma}}(\ell) &=& \int \frac{{\rm d}\chi}{r^2(\chi)}
W_{\rm ps}(\chi) W_{\kappa}(\chi) 
\nonumber \\
&&
\,\,\,\,\,
\,\,\,\,\,
\times
b_{\rm ps}(z) P_{L}\left(k_{\ell}=\frac{\ell+1/2}{r(\chi)}, z(\chi)\right), \label{eq:model_cross_power_stack} \\
W_{\rm ps} &=& {\cal C} \chi^2 \int_{L_{\rm min}(z)} {\rm d}L_{\gamma} \Phi_{\rm ps}(L_{\gamma},z),
\label{eq:window_ps}
\eeqa
where $\Phi_{\rm ps}(L_{\gamma},z)$ is the luminosity function of point sources
and the window function $W_{\rm ps}$ is normalized to $\int {\rm d}\chi\, W_{\rm ps} = 1$.
We simply assume that all the resolved sources in our ROIs are blazars.
The lower limit in the integral in Eq.~(\ref{eq:window_ps}) is given by the flux limit in the $\gamma$-ray observations.
To compute $L_{\rm min}$, we assume the $\gamma$-ray energy spectrum of blazars (see Eq.~7) and 
set the flux limit to be $2\times10^{-9}\, {\rm cm}^{-2}\, {\rm s}^{-1}$ above 100 MeV.
}

\subsubsection*{Result}

\begin{figure}
\begin{center}
        \includegraphics[clip, width=0.9\columnwidth]
       {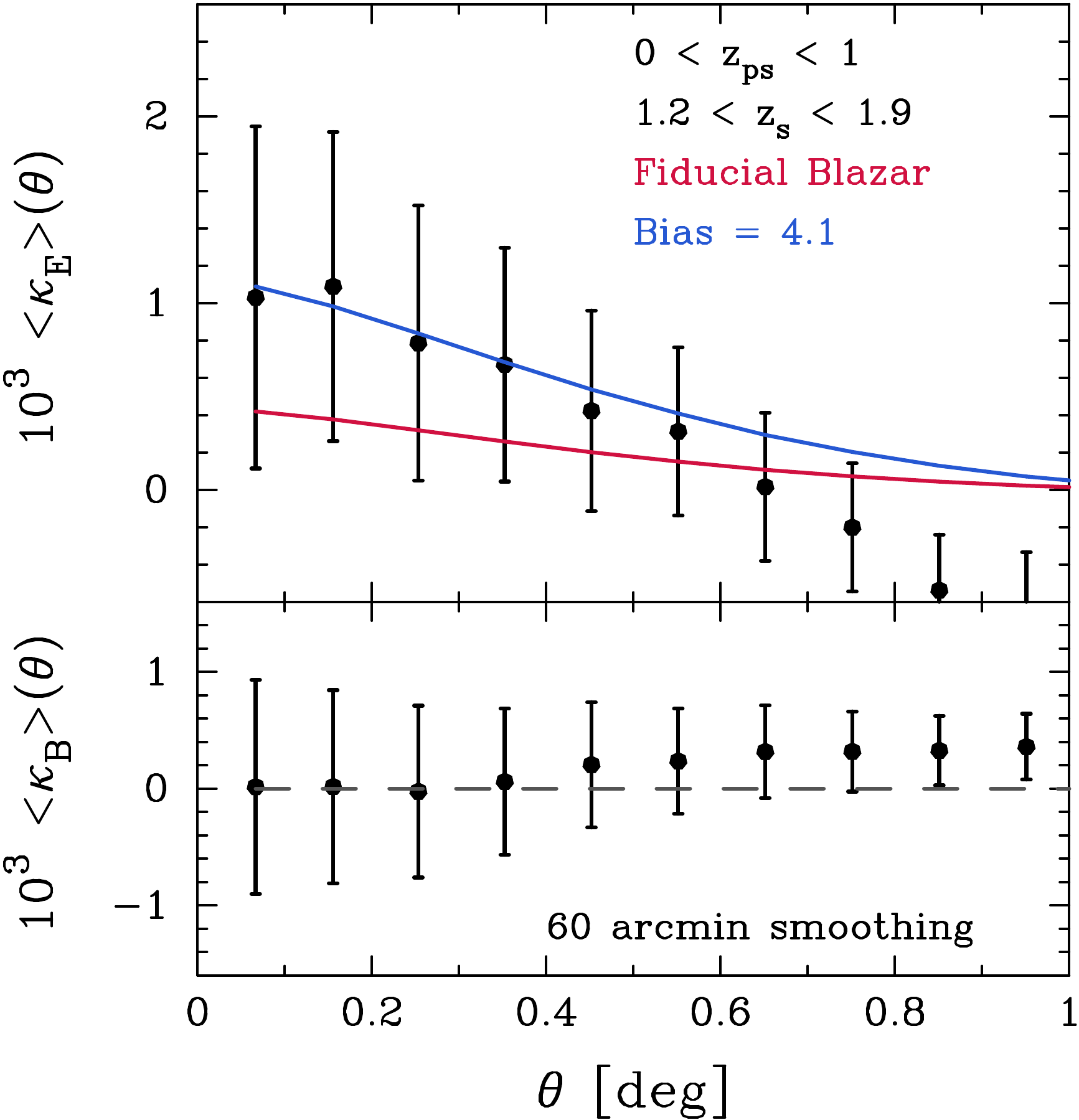}
     \caption{
     \label{fig:stacked_kappa}
     \change{
	 The measurement of the average convergence profile around $\gamma$-ray point sources in the HSCS16A fields.
     The upper panel shows the stacked profile of lensing convergence $\langle\kappa_E\rangle$
     as a function of separation angle from a point source, $\theta$.
     To reconstruct $\kappa_E$ from galaxy shapes, we impose a photometric redshift selection of 
     $1.2<z_{\rm photo}<1.9$ and set a Gaussian smoothing scale of 60 arcmin.
     To increase the detectability of lensing effects, 
     we also consider foreground $\gamma$-ray sources by using their redshifts $z_{\rm ps}$ smaller than 1.
     In the upper panel, the red line shows the expected convergence profile from our fiducial model of blazar clustering,
     while the blue line is for a simple model with a large blazar bias of $4.1$.
     Note that a bias of $4.1$ provides the minimum chi-squared value for $\langle\kappa_E\rangle$ at $\theta<0.6\, {\rm deg}$.
     The lower panel plots the B-mode convergence $\langle \kappa_B \rangle$, which is expected 
     to be zero in the absence of significant systematic effects in the lensing survey.
     As shown in the lower panel, we find a non-zero 
     $\kappa_B$ at $\theta\simgt0.6\, {\rm deg}$,
     indicating the large-scale $\kappa_E$ signals are not reliable for quantitative analyses. 
     }
  }
    \end{center}
\end{figure}

\change{
Figure \ref{fig:stacked_kappa} shows the stacked convergence profile around resolved $\gamma$-ray sources.
We reconstruct the lensing convergence with a Gaussian smoothing scale of $60\, {\rm arcmin}$; 
the large smoothing scale assures that our measurements are insensitive to non-linear clustering of point sources.
To select the background source galaxies robustly in the lensing analysis, and to reduce the possible contamination by high-z $\gamma$-ray sources,
we select the HSC galaxies by photometric redshifts to be in the range of $1.2<z_{\rm photo}<1.9$. We also use only $\gamma$-ray point sources at $z<1$.
After the redshift selection, we find 118 $\gamma$-ray sources available for our analysis.
When performing the stack analysis, we set the angular bin width to be $\Delta\theta=0.1\, {\rm deg}$.
The number of random points is set to be 100 times as large as the number of point sources. 
To estimate the statistical uncertainty, we utilize 200 mock HSC catalogs.

The bottom panel in figure~\ref{fig:stacked_kappa} shows non-zero B-mode convergence at large angular scales of $\theta>0.6\, {\rm deg}$.
It indicates that the current data do not allow us to derive 
$\langle\kappa_E\rangle$ accurately,
and thus we simply ignore the large-scale signals in the following discussions.
Similarly to the method in Section~\ref{subsec:cross_corr}, we compute the detection significance of the stacked convergence profile
(also see Eq.~\ref{eq:chi2}).
We find that $\chi_{0}^2=3.75$ for $\langle \kappa_E \rangle$
and $2.89$ for $\langle \kappa_E \rangle$ with 6 degrees of freedom.
Hence, the current measured signal is still consistent with null detection\footnote{
We have also confirmed that the stacked signal is still consistent with a null detection
when setting the smoothing scale to $30$ arcmin.
In this case, the detection significance is found to be $\chi_{0}^2=6.03$ for $\langle \kappa_E \rangle$
and $3.19$ for $\langle \kappa_B \rangle$ with 6 degrees of freedom.
At face value, a bias of $\sim4.8$ can provide a reasonable fit to the observed 
$\langle \kappa_E \rangle$ in this case.
}.
Although the statistical errors are large, we find that $b_{\rm ps}=4.1$ in Eq.~(\ref{eq:model_cross_power_stack}) 
provides a reasonable fit to the observed $\langle \kappa_E \rangle$, as shown by the solid line in the Figure. Similarly to our finding in Section \ref{subsec:implication}, 
we observe that the stacked lensing analysis also implies a large bias of blazars in our ROIs.
}

\section{\label{sec:con}CONCLUSION AND DISCUSSION}

In this paper, we have calculated the cross correlation of the EGB (i.e., the whole 
extragalactic $\gamma$-ray radiation including resolved sources) 
and the cosmic matter density at $z\simlt1$ for the first time.
One may naively expect a strong correlation between the two 
if the EGB is mostly contributed by known astrophysical sources \cite{Ajello:2015mfa}.
The cross correlation at degree angular scales is found to be consistent 
with a null detection given the current statistical uncertainties. 
The result still allows us to constrain possible relations
between the dominant source population of the EGB and the
underlying mass distribution.
We have tested and confirmed that our measurement is robust 
in terms of residual systematic effects in both the $\gamma$-ray
and lensing analyses.

Provided that the astrophysical models in Section~\ref{subsec:EGB}
can reproduce the observed EGB spectrum well,
we argue that the correlation of blazars 
and the large-scale matter density dominates the measured cross correlation
at degree scales.
This allows us to derive an estimate of the blazar bias with 
respect to the underlying matter density at $z\sim 0.5$.
The clustering bias can be used to infer
the host environment of the blazar population in the hierarchical
structure formation model, and also to examine the unified AGN scenario 
\cite{Urry:1995mg}.
\change{
Although our cross-correlation measurements are still consistent with a null detection,
we find that a large bias of $\sim 4$ provides a reasonable explanation for the 
EGB-lensing correlations.
The blazar bias of $\sim4$ is also favored in order to explain the stacked lensing convergence profile around the resolved $\gamma$-ray sources. 
Our measurements allow to constrain a constant blazar bias.
Interestingly, our result is consistent with other observations of AGN bias selected 
at different wavelengths (Figure~\ref{fig:bias_constraint}).
}

\change{
The HSC final data with a sky coverage of 1,400 squared degrees 
will allow us to measure the clustering bias of blazars more accurately.
}
Assuming that the statistical uncertainty of the cross-correlation
is reduced with survey area,
\change{
we should be able to detect the correlation between blazars and
the matter density at a $\sim 3 \sigma$ level
if the $\gamma$-ray sources are strongly clustered as indicated in the present paper.
The cross correlation of the EGB and the HSC final data 
will offer invaluable information on the sources that give a dominant contribution 
to the EGB at $z > 0.5$, and provide a stringent statistical 
test of the unified AGN scenario at $\gamma$-ray energy scales.
}
If the blazar bias is determined accurately, one can develop 
realistic models of the blazar population in the context 
of structure formation in a similar manner 
to that applied to SDSS galaxies \cite{vandenBosch:2006iq}.

Identifying the sources of the EGB remains a major goal for gamma-ray
astrophysics.
We need multiple information  
to reveal possible contributions from 
star-forming galaxies, radio galaxies, and diffuse processes.
Joint measurements of 
multiple cross-correlation functions will be the key to this end. 
It is suggested that the cross-correlation of local galaxies and the cosmic infrared 
background helps to infer the  
star-forming activity in galaxies over a wider 
range of redshifts \cite{Xia:2015wka, Feng:2016fkl, Cuoco:2017bpv}, 
while observations of galaxy clusters, radio galaxies,
and the CMB lensing effect will tighten
the constraints on the contribution from misaligned AGN
\cite{Fornengo:2014cya, Branchini:2016glc}.
Furthermore, a tomographic approach using
$\gamma$-ray energies and redshifts of LSS tracers 
will be viable with upcoming galaxy surveys 
\cite{Ando:2014aoa, Camera:2014rja}.
Ultimately, by putting all these information together, we will be able to understand the contributions to the EGB and perform stringent 
searches on the particle nature of dark matter. This 
{\it cosmological} probe will be highly complementary to searches using $\gamma$-ray measurements of local galaxies \cite{Hooper:2011ti, Ackermann:2013yva}.

\acknowledgements
\change{The authors thank an anonymous referee 
for careful reading and suggestion to improve the article.} 
We also thank Yutaka Komiyama for providing useful comments on the manuscript. 
The work of SH is supported by the U.S.~Department
of Energy under award number
DE-SC0018327. 
NY acknowledges financial support from JST CREST (JPMJCR1414).
Numerical computations presented in this paper were in part carried out on the general-purpose PC farm at Center for Computational Astrophysics,
CfCA, of National Astronomical Observatory of Japan.

The Hyper Suprime-Cam (HSC) collaboration includes the astronomical communities of Japan and Taiwan, and Princeton University.  The HSC instrumentation and software were developed by the National Astronomical Observatory of Japan (NAOJ), the Kavli Institute for the Physics and Mathematics of the Universe (Kavli IPMU), the University of Tokyo, the High Energy Accelerator Research Organization (KEK), the Academia Sinica Institute for Astronomy and Astrophysics in Taiwan (ASIAA), and Princeton University.  Funding was contributed by the FIRST program from Japanese Cabinet Office, the Ministry of Education, Culture, Sports, Science and Technology (MEXT), the Japan Society for the Promotion of Science (JSPS),  Japan Science and Technology Agency  (JST),  the Toray Science  Foundation, NAOJ, Kavli IPMU, KEK, ASIAA,  and Princeton University.

The Pan-STARRS1 Surveys (PS1) have been made possible through contributions of the Institute for Astronomy, the University of Hawaii, the Pan-STARRS Project Office, the Max-Planck Society and its participating institutes, the Max Planck Institute for Astronomy, Heidelberg and the Max Planck Institute for Extraterrestrial Physics, Garching, The Johns Hopkins University, Durham University, the University of Edinburgh, Queen's University Belfast, the Harvard-Smithsonian Center for Astrophysics, the Las Cumbres Observatory Global Telescope Network Incorporated, the National Central University of Taiwan, the Space Telescope Science Institute, the National Aeronautics and Space Administration under Grant No. NNX08AR22G issued through the Planetary Science Division of the NASA Science Mission Directorate, the National Science Foundation under Grant No. AST-1238877, the University of Maryland, and Eotvos Lorand University (ELTE).

This paper makes use of software developed for the Large Synoptic Survey Telescope. We thank the LSST Project for making their code available as free software at \verb|http://dm.lsst.org|.

Based [in part] on data collected at the Subaru Telescope and retrieved from the HSC data archive system, which is operated by the Subaru Telescope and Astronomy Data Center at National Astronomical Observatory of Japan.

\appendix

\section{\emph{FERMI}-LAT ANALYSIS METHODS}\label{sec:gammaanalysis}

The analysis of the full $20^\circ \times 20^\circ$ \emph{Fermi}-LAT data sets encompassing each of the HSC regions was divided into four smaller regions, each of which was modeled by a diffuse $\gamma$-ray background and $\gamma$-ray point source model. Namely, we divided each $20^\circ \times 20^\circ$ region into four $12^\circ \times 12^\circ$ contiguous patches with overlapping boundaries for which we started the fit with a sky model that includes all point-like and extended LAT sources listed in the 3FGL~\cite{3FGL} catalog as well as with models for the Galactic diffuse and isotropic emission. In our baseline model the Galactic diffuse emission was modeled by the standard LAT diffuse emission model {\tt gll\_iem\_v06.fits}, and as a proxy for the residual background and extragalactic $\gamma$-ray radiation spectra we used a single isotropic component with the spectral shape given by model {\tt iso\_P8R2\_ULTRACLEANVETO\_V6\_v06.txt}\footnote{http://fermi.gsfc.nasa.gov/ssc/}.

We used the \emph{Fermipy} Python package\footnote{
  \url{http://fermipy.readthedocs.io/en/latest/}.} in conjunction
with standard \emph{Fermi Science Tools} to fit and characterize the sources included in the sky models corresponding to each region of interest (ROI). In each ROI we bin the LAT data in 8 energy bins per decade and with a spatial pixel size of $0.1^{\circ}$.  Each ROI is analyzed separately; however, in order to perform our cross correlation analysis in the full HSC regions we merge the patches from the analyses of the different ROIs.

Using the initial baseline model mentioned above, the first step of our method is to find the best spectral parameters for all free sources within our $12^\circ \times 12^\circ$ ROIs. To obtain convergence, all the fits were performed hierarchically; freeing first the normalization of the sources with the highest intensities followed by the lower ones within the ROIs. The fitting consecutively restarts from the updated best-fit models and repeats the same procedure this time for the spectral shape parameters. Due to the $\sim 7$ years of LAT data used in our analysis compared to the 4 years of the 3FGL~\cite{3FGL} catalog, we expected to find new $\gamma$-ray point source candidates in our ROIs. The significance of each new point source candidate was evaluated using the test statistic ${\rm TS}=2(\ln \mathcal{L}_1 - \ln \mathcal{L}_0)$, where $\mathcal{L}_0$ and $\mathcal{L}_1$ are the likelihoods of the background (null hypothesis) and the hypothesis being tested (alternative hypothesis: background plus source). All point source candidates that were found to have a TS~$\geq 25$ were included to our sky model of the respective ROI. The new \emph{Fermipy} package includes special routines to perform automated point source searches, and in this study we followed the standard method recommended in the \emph{Fermipy} documentation. We then used this package to refine the positions and the spectral parameters of the new point source candidates. Since the set of four $12^\circ \times 12^\circ$ patches making up every $20^\circ \times 20^\circ$ HSC ROI have some overlapping regions at the boundaries, we remove any duplicate point source candidates found in more than one patch. In this study we detected 48 new point source candidates with TS~$\geq 25$ in the combined the HSC ROIs. The positions and spatial overlaps with the Roma-BZCAT~\cite{Massaro:2009} Blazar catalog are shown in Table~\ref{tab:ptsrcs}. All the characteristics of the new point source candidates are provided in FITS files with this article.

The EGB photons in the ROIs were obtained by subtracting the best-fit Galactic diffuse emission model from the photon counts maps. We note that the EGB images obtained in this way could still contain some isotropic detector backgrounds. However, our analysis is able to reproduce well the EGB $\gamma$-rays derived by the \emph{Fermi}-LAT team (see, e.g.,  Ref.~\cite{Shirasaki:2016kol} for an example of this method). This shows 
that most of the detector cosmic-ray induced backgrounds are safely removed by our conservative photon selection filters.

We estimated the systematic uncertainties in the Galactic diffuse emission model in a similar fashion to that explored in the Fermi collaboration paper on the EGB~\cite{Ackermann:2014usa}.  Namely, we reran our pipeline using the alternative foreground Models A, B and C as described in the appendix of Ref.~\cite{Ackermann:2014usa}.  Such foreground models encompass a very generous range of the systematics associated with this kind of analysis, and provides a test in rigor comparable to that performed by the \emph{Fermi} team.

\newpage 
\clearpage
\begin{longtable*}[t]{C{0.3\textwidth}C{0.1\textwidth}C{0.1\textwidth}C{0.3\textwidth}p{2cm}}
\caption{
 New Point Source Candidates detected in the HSC ROIs.  
	\label{tab:ptsrcs}
    }\\  
	\hline\hline	Name &   $l$    &   $b$     &    Association     & TS     \\
	     &   [deg]  &   [deg]   & or spatial overlap &       \\ \hline
\endfirsthead
\endhead
PSJ0922.8+0432	&	227.74	&	35.48	&		&	44	\\
PSJ0932.7+1040	&	222.40	&	40.57	&	5BZBJ0932+1042	&	28	\\
PSJ0930.7+0031	&	233.24	&	35.02	&	5BZQJ0930+0034	&	76	\\
PSJ0833.6-0455	&	229.77	&	20.07	&		&	35	\\
PSJ0848.0-0704	&	233.77	&	21.97	&	5BZBJ0847-0703	&	101	\\
PSJ0819.5-0755	&	230.58	&	15.53	&	5BZBJ0819-0756	&	36	\\\hline
									
PSJ1420.2+0612	&	352.20	&	60.28	&	 5BZBJ1420+0614 	&	37	\\
PSJ1410.1+0202	&	343.20	&	58.62	&	5BZBJ1410+0203	&	139	\\
PSJ1356.6+0237	&	338.22	&	60.92	&		&	48	\\
PSJ1443.5-0959	&	342.98	&	43.99	&		&	34	\\
PSJ1405.1-0642	&	333.41	&	51.74	&		&	30	\\
PSJ1359.1-0658	&	331.03	&	52.17	&		&	29	\\
PSJ1401.0-0914	&	330.23	&	49.90	&	5BZQJ1401-0916	&	61	\\
PSJ1425.3-0119	&	345.15	&	53.66	&	5BZBJ1425-0118 	&	33	\\\hline
									
PSJ1644.2+4544	&	71.33	&	40.84	&	5BZGJ1644+4546	&	42	\\
PSJ1607.8+4947	&	77.85	&	46.40	&		&	36	\\
PSJ1604.2+5008	&	78.59	&	46.91	&	5BZBJ1603+5009	&	54	\\
									
PSJ1607.8+4950	&	77.90	&	46.39	&		&	42	\\
PSJ1536.0+3744	&	60.58	&	54.02	&	5BZUJ1536+3742	&	253	\\
PSJ1529.5+3813	&	61.70	&	55.23	&	5BZBJ1529+3812	&	29	\\
PSJ1602.1+3328	&	53.77	&	48.71	&	 5BZUJ1602+3326  	&	33	\\\hline
									
PSJ2249.9+0451	&	75.76	&	-46.59	&		&	31	\\
PSJ2226.5+0207	&	67.11	&	-44.48	&		&	28	\\
PSJ2235.2-0631	&	59.21	&	-51.68	&		&	38	\\
PSJ2301.0-0157	&	71.81	&	-53.49	&	5BZQJ2301-0158	&	136	\\
PSJ2225.7-0803	&	55.01	&	-50.63	&		&	45	\\
PSJ2156.1-0036	&	57.71	&	-40.32	&	5BZUJ2156-0037	&	127	\\
PSJ2211.0-0004	&	61.32	&	-42.96	&	5BZBJ2211-0003	&	35	\\
PSJ2148.0-0734	&	48.46	&	-42.42	&	5BZBJ2148-0733	&	80	\\\hline
									
PSJ1219.7+0446	&	282.89	&	66.39	&	5BZBJ1219+0446	&	47	\\
PSJ1219.7+0547	&	282.00	&	67.39	&		&	37	\\
PSJ1215.1+0734	&	277.57	&	68.63	&	5BZGJ1215+0732	&	53	\\
PSJ1216.1+0931	&	276.01	&	70.52	&	5BZGJ1216+0929	&	58	\\
PSJ1216.0-0243	&	285.59	&	58.96	&	5BZBJ1216-0243	&	36	\\
PSJ1207.6-0104	&	280.65	&	59.89	&	5BZQJ1207-0106	&	69	\\
PSJ1136.0-0426	&	270.09	&	53.55	&	5BZQJ1135-0428	&	45	\\
PSJ1129.3-0530	&	268.53	&	51.82	&		&	59	\\
PSJ1131.0-0945	&	272.35	&	48.29	&		&	45	\\\hline
									
PSJ0237.3+0206	&	168.27	&	-51.19	&		&	30	\\
PSJ0239.8+0417	&	166.90	&	-49.10	&	5BZQJ0239+0416	&	95	\\
PSJ0245.0-0257	&	176.18	&	-53.65	&		&	55	\\
PSJ0208.6-0045	&	161.24	&	-57.79	&	5BZBJ0208-0047	&	38	\\
PSJ0220.9-0841	&	176.01	&	-61.94	&	5BZBJ0220-0842 	&	38	\\
PSJ0226.6-0553	&	174.07	&	-58.96	&		&	63	\\
PSJ0241.0-0506	&	177.65	&	-55.86	&	5BZBJ0240-0504 	&	51	\\
PSJ0205.8-0955	&	172.03	&	-65.42	&		&	33	\\
PSJ0140.7-0758	&	156.59	&	-67.57	&	5BZBJ0140-0758	&	42	\\
PSJ0142.6-0543	&	154.87	&	-65.36	&	5BZBJ0142-0544	&	35	\\\hline

\multicolumn{5}{p{17cm}}{New point source candidates in the HSC ROIs found with a $TS\geq 25$ in $\sim 7$ years of \emph{Fermi}-LAT data. This table is also provided as a FITS file in the online material. The horizontal dividers separate the new point source candidates in the GAMA09H, GAMA15H, HECTOMAP, VVDS, WIDE12H and XMM patches, from top to bottom. The fourth column displays associations or spatial overlaps with tentative multi-wavelength counterparts. For this purpose we use the Roma-BZCAT blazar catalog~\cite{Massaro:2009}. To identify possible multi-wavelength counterparts to the new point source candidates we searched in the seed locations within the 68\% containment of the point spread function for one of our high-energy bands, which comes to $\sim 0.2^\circ$.}

\end{longtable*}


\section{CROSS-CORRELATION WITH ISOTROPIC GAMMA-RAY COMPONENTS 
AND CONSTRAINTS ON DARK MATTER ANNIHILATION}

\begin{figure*}
\begin{center}
       \includegraphics[clip, width=0.85\columnwidth, bb=0 0 524 507]
       {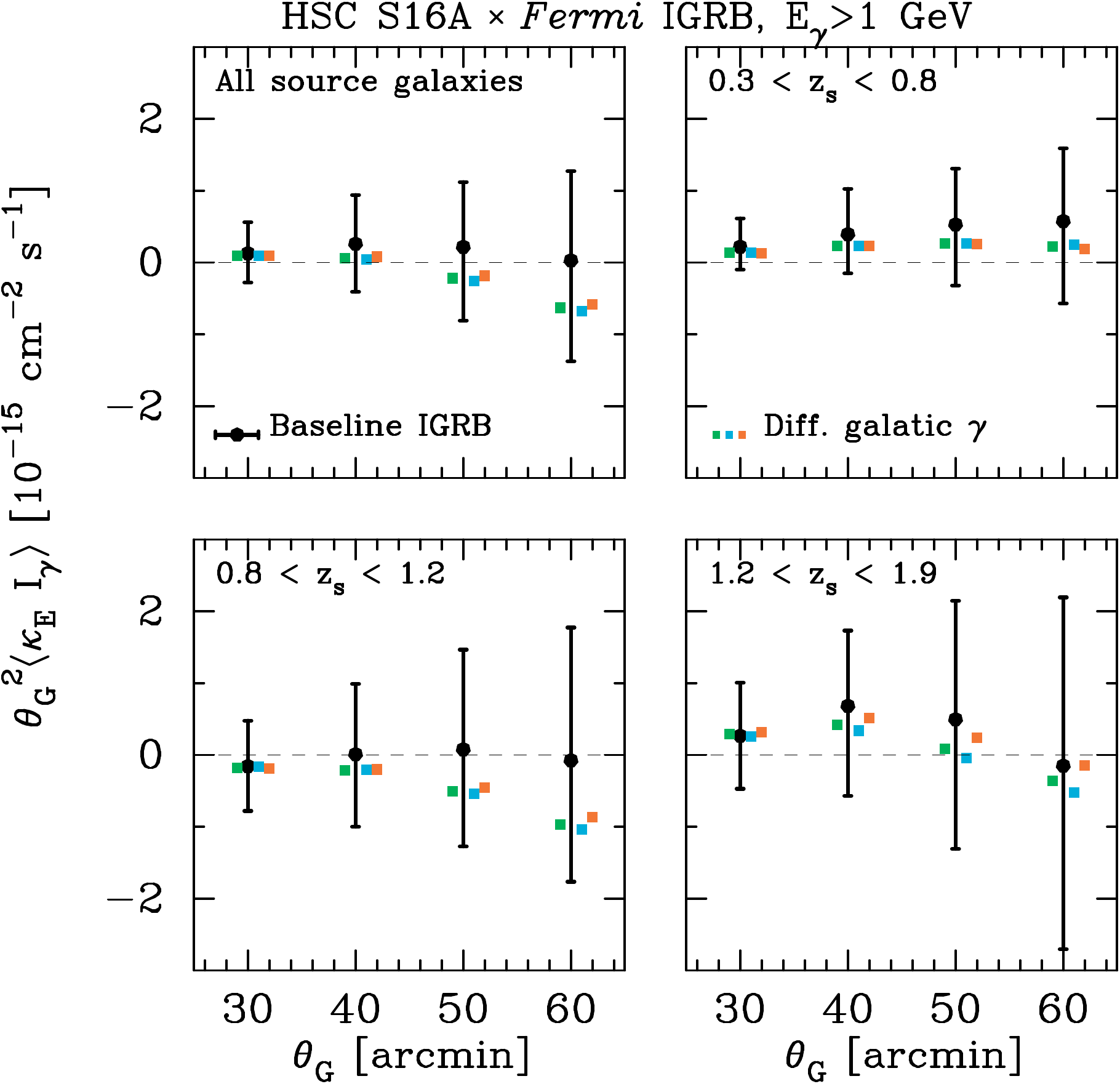}
       \includegraphics[clip, width=0.9\columnwidth, bb=0 0 523 466]
       {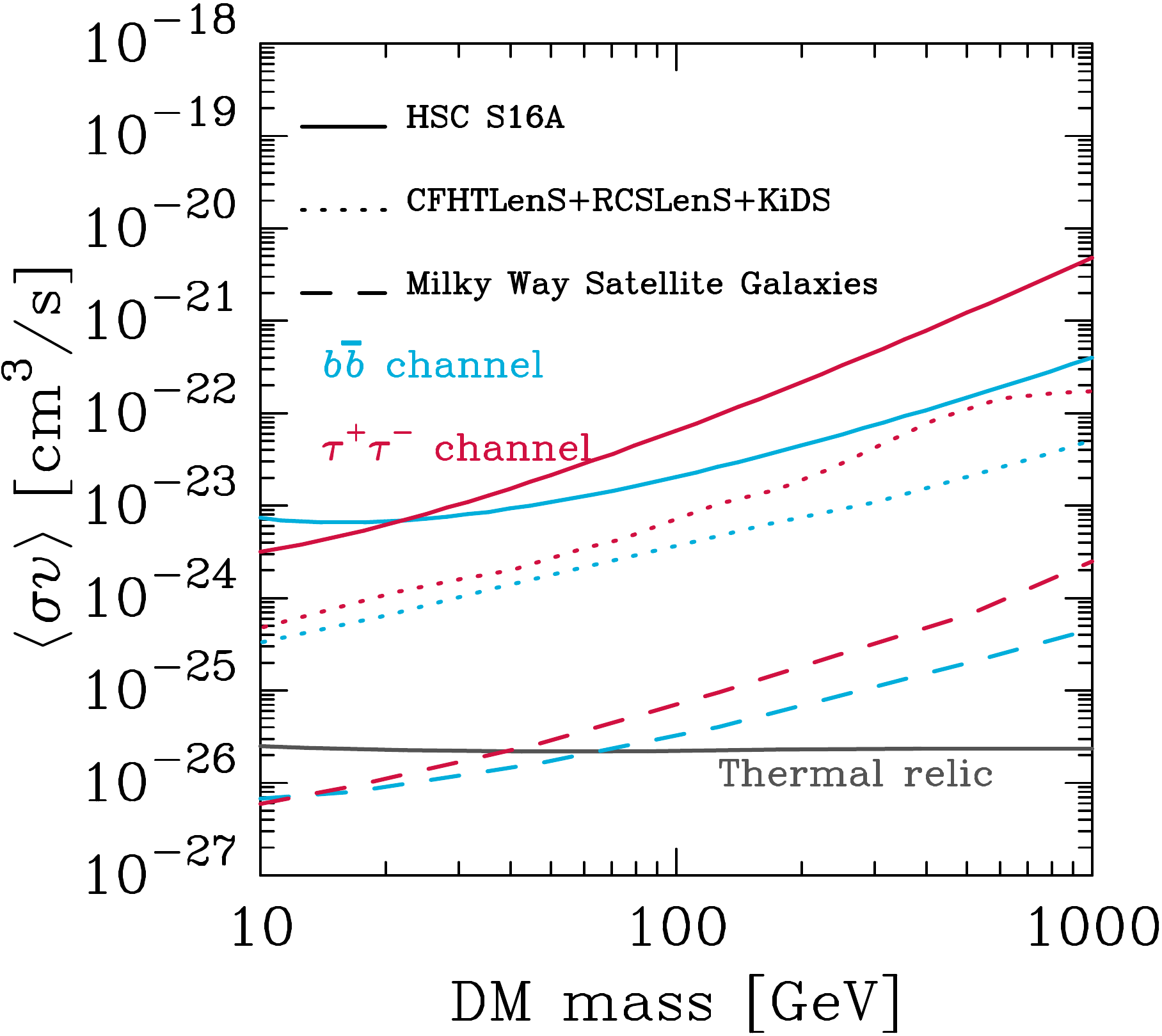}
     \caption{
     \label{fig:cross_corr_IGRB_kappa}
     {\it Left}: 
     The cross-correlation of the 
     IGRB and smoothed lensing convergence. Each panel shows the result
     obtained with different source-galaxy selections by photometric
     redshifts in the lensing analysis. The black points with error 
     bars represent results with our baseline IGRB maps, while colored 
     points are for the cases when we use different galactic $\gamma$-ray 
     templates in the Fermi-LAT analysis pipeline.
     No significant correlations are found in all cases. Possible 
     systematic effects by imperfect subtraction of galactic $\gamma$ 
     rays can be safely ignored in our analysis.
     {\it Right}:
     95\% confidence level upper limits on the DM annihilation cross 
     section as a function of DM particle mass. The different colors 
     correspond to two different annihilation channels ($b\bar{b}$ 
     and $\tau^{+}\tau^{-}$).
     The solid lines present constraints from our cross correlation 
     measurements using the HSCS16A data, while the dotted lines are 
     for similar cross-correlation analysis with wider survey area 
     coverage \cite{Troster:2016sgf}, and the dashed lines show the 
     $\gamma$-ray emission constraints from local dwarf galaxies 
     \cite{Ackermann:2013yva}.
     The gray line shows the canonical cross section expected for 
     thermal relic DM \cite{2012PhRvD..86b3506S}.     
  } 
    \end{center}
\end{figure*}

In this appendix, we present the cross-correlation analysis of the lensing convergence and 
the isotropic $\gamma$-ray background (IGRB) including
the unresolved components in Fermi-LAT data.

The lensing convergence $\kappa_E (\theta)$ is equivalent to the cosmic 
surface mass density with a weight function along line-of-sight 
direction $\theta$. The details of the reconstruction algorithm of $\kappa_E$ 
from the HSCS16A galaxy-imaging data are given in Section~\ref{subsec:reconst_kappa}.
We use the cross-correlation estimator in Eq.~(\ref{eq:cross_corr_est}) 
to the IGRB measured by Fermi-LAT and the lensing convergence 
from HSCS16A. As in Section~\ref{subsec:est_cross_corr},
we evaluate the statistical uncertainty of the cross correlation 
with 200 HSCS16A mock catalogs and use the inverse-weighed method
to combine the cross-correlation signals for individual HSCS16A patches together.
In addition to the normal masks as described in Section~\ref{subsec:est_cross_corr},
we also place circular masks around Fermi-LAT $\gamma$-ray point 
sources, with a radius of 1 deg. The circular masks are introduced 
in order to minimize possible contaminants from residual $\gamma$- 
rays from resolved bright sources.

\begin{table}[!t]
\begin{center}
\begin{tabular}{|c|c|c|}
\tableline
Tomographic bins & 
$\langle \kappa_E I_\gamma \rangle$  &
$\langle \kappa_B I_\gamma \rangle$  \\ \hline \hline
No selection in $z_{\rm photo}$ 
& 1.24 (4) & 5.50 (4) \\
$0.3<z_{\rm photo}<0.8$ 
& 0.72 (4) & 6.57 (4) \\
$0.8<z_{\rm photo}<1.2$ 
& 1.13 (4) & 5.87 (4) \\
$1.2<z_{\rm photo}<1.9$ 
& 4.24 (4) &  0.99 (4) \\ \tableline
Combined 
& 7.06 (16) & 16.6 (16) \\ \tableline
\end{tabular}
\caption{
\label{tab:chi2_IGRB} 
Summary of the significance of 
our IGRB-$\kappa_E$ cross-correlation measurements.
Second and third columns list the $\chi_{0}^2$ values defined in 
Eq.~(\ref{eq:chi2}), and the number with bracket shows the degree of freedom in the analysis. 
}
\end{center}
\end{table}

The left panels in Figure~\ref{fig:cross_corr_IGRB_kappa}
summarize our cross correlation measurements of the IGRB and 
$\kappa_E$. We work with four different source-galaxy 
selections in the lensing analysis. Regardless of the choice 
of smoothing scales and galaxy selections, we find that
the IGRB-$\kappa_E$ cross-correlation is consistent with 
a null detection. We also confirm that the cross-correlation 
measurements of the IRGB and the B-mode convergence is 
consistent with a null detection, suggesting that our lensing 
analysis is not compromised by systematic errors. 
Table~\ref{tab:chi2_IGRB} summarizes the significance of
our results using the definition in Eq.~(\ref{eq:chi2}).

We can derived constraints on dark matter (DM) annihilation 
directly using the cosmic matter density distributions.
Using the formulation as described in Section~\ref{subsec:form}, 
we compute the expected correlation originating from DM annihilation. 
We adopt the model of cross power-spectrum between the matter 
density and $\gamma$-rays from DM annihilation as in Ref.~\cite{Shirasaki:2016kol} (see their Eq.~[14]).
We then consider a conservative model of substructures in DM 
halos \cite{Sanchez-Conde:2013yxa}.
Following a similar likelihood analysis as in 
Section~\ref{subsec:implication}, 
we can constrain the two parameters in DM annihilation,
DM particle mass and the annihilation cross-section 
$\langle\sigma v\rangle$, for a given annihilation channel.
The right panel in Figure~\ref{fig:cross_corr_IGRB_kappa} 
shows our cosmological constraints on DM annihilation.
For comparison, we also present the constraints from 
similar analyses with shallower but 
wider galaxy survey data \cite{Troster:2016sgf},
and from analyses of the Milky-Way satellite galaxies \cite{Ackermann:2013yva}.
Since our measurements in the HSCS16A cover a smaller sky 
area than Ref.~\cite{Troster:2016sgf},
our constraints are found to be weaker.
Also it is unlikely for us to be able to place
constraint on DM annihilation at a similar level as the
local measurements \cite{Ackermann:2013yva},
even if we use the HSC final data release, because the
increased sky coverage can improve the statistical 
uncertainty only by a factor of $\sim3-4$.

There are two reasons why our analysis does not provide 
tight constraints on DM annihilation:
\begin{enumerate}
\item
Our cross-correlation analysis uses 
linear-scale clustering alone (see Section~\ref{subsec:form})
that contains less information on the internal mass
distribution of DM halos.

\item
HSC is suited for measuring the matter 
distributions in the distant universe. The median source 
redshift in HSC is larger than those in other data in the 
literature.
Cross-correlation of the IGRB with objects or the matter
density at lower redshifts can be more optimal for indirect 
detection of DM annihilation
(also see Ref.~\cite{Shirasaki:2015nqp} for optimal 
targets in large-scale structure for indirect detection 
of DM annihilation).
\end{enumerate}

Overall, the cross-correlation measurement presented 
in this paper is a more promising tool to explore expected correlations 
of astrophysical $\gamma$-ray sources with the cosmic 
matter density.

\bibliography{ref_prd}
\end{document}